\documentclass[aps,prl,twocolumn,showpacs,superscriptaddress]{revtex4-2}

\usepackage{amsmath}
\usepackage{graphicx}
\usepackage{lmodern}
\usepackage{amsmath}

\usepackage{color}
\usepackage{amssymb}
\usepackage{bm}
\usepackage{braket}
\usepackage{mathtools}
\usepackage{slashed}

\usepackage{diagbox}

\usepackage[dvipsnames]{xcolor}
\usepackage[colorlinks=true]{hyperref}
\hypersetup{
    colorlinks=true,
    linkcolor=blue,
    citecolor=blue,
    urlcolor=blue
}

\makeatletter
\def\maketitle{
\@author@finish
\title@column\titleblock@produce
\suppressfloats[t]}
\makeatother
\begin{document}

\title {Exactly solvable pair-density wave  in topological flat bands from magnetic translation symmetries}
\date{\today}
\author{Zhengzhi Wu}
\affiliation{Rudolf Peierls Centre for Theoretical Physics, Parks Road, Oxford, OX1 3PU, UK}
\author{Zhou-Quan Wan}
\affiliation{Center for Computational Quantum Physics, Flatiron Institute, New York, NY 10010, USA}
\author{Steven H. Simon}
\affiliation{Rudolf Peierls Centre for Theoretical Physics, Parks Road, Oxford, OX1 3PU, UK}

\begin{abstract}
Pair-density wave (PDW) superconductors are exotic phases in which Cooper pairs carry finite center-of-mass momentum. Despite a variety of theoretical and experimental reports on PDW states, exact PDW ground states in topological bands have remained elusive.  Here we construct exactly solvable models with PDW ground states in topological flat bands using a generalized version of the recently proposed quantum geometric nesting (QGN) framework. Our construction broadly applies to systems with non-commuting magnetic translation symmetries (MTS) and time-reversal symmetry, exemplified by the time-reversal invariant version of the Kapit-Mueller model with two ideal flat Chern bands with opposite Chern number. 
Our construction thus provides a  platform for further studies of  band topology and quantum geometry in PDW superconductors.
\end{abstract}
\maketitle

{\bf Introduction:} Pair-density wave (PDW) superconductors represent a novel class of superconducting phases in which the order parameter oscillates periodically in space, such that its spatial average vanishes~\cite{annurev}. Not limited to the early Fulde–Ferrell–Larkin–Ovchinnikov (FFLO) states~\cite{PhysRev.135.A550,LO}—where finite-momentum pairing arises under an external magnetic field that shifts the maximum of the pairing susceptibility from zero to finite momentum—a PDW state can also emerge in systems that preserve time-reversal symmetry (TRS). It is generally believed, however, that PDW order cannot develop in time-reversal-symmetric systems in the weak-coupling limit \cite{annurev}. There has been growing theoretical interest \cite{PhysRevLett.99.127003,Berg_2009,PhysRevX.4.031017,PhysRevLett.105.146403,PhysRevB.82.041102,PhysRevB.85.035104,PhysRevB.101.165133,PhysRevB.106.045103,Yang_2009,Berg2009,PhysRevLett.114.237001,PhysRevLett.118.166802,PhysRevB.86.214514,PhysRevB.89.165126,PhysRevLett.114.197001,PhysRevB.94.180506,PhysRevB.98.224501,Zhou2022,PhysRevLett.129.167001,PhysRevB.108.035135,PhysRevB.107.224516,PhysRevLett.130.026001,PhysRevLett.130.126001,Setty2023,PhysRevLett.131.016002,PhysRevB.107.045122,PhysRevLett.131.026601,PhysRevX.9.021047,PhysRevB.109.L121101,doi:10.1126/sciadv.aat4698,PhysRevLett.122.167001,Peng2021,Peng_2021,PhysRevB.107.214504,PhysRevLett.133.176501,10.3389/femat.2023.1323404,PhysRevB.108.L201110,Wang2025,PhysRevLett.125.167001,PhysRevX.14.041004,PhysRevB.108.174506,PhysRevB.101.054506,PhysRevB.110.094515,Zheng2025} and accumulating experimental evidence \cite{PhysRevB.106.174510,lee2023,PhysRevB.83.104506,PhysRevB.85.134513,Hamidian2016,Ruan2018,doi:10.1126/science.aat1773,PhysRevX.11.011007,Jiao2020,Chen2021,PhysRevX.14.021025,Aishwarya2023,Gu2023,Zhao2023,Liu2023} for PDW states. The study of PDW has uncovered a wealth of novel physics, including exotic quantum critical points with emergent spacetime supersymmetry \cite{PhysRevLett.114.237001,PhysRevLett.118.166802}, and a variety of vestigial orders, such as charge-4e or -6e superconductors \cite{Berg2009,PhysRevX.14.021025,Zhou2022,PhysRevB.111.054508}.  However, establishing a PDW state in an unbiased theoretical framework is generally challenging, and a general microscopic mechanism for PDW formation is still unknown \cite{annurev}. 

An even more intriguing question concerns PDW states, or more broadly, superconductivity, in topological bands, owing to the nontrivial implications of band topology for superconducting phenomena (e.g. the lower bound of superfluid stiffness~\cite{Peotta2015,PhysRevLett.124.167002,PhysRevLett.123.237002,PhysRevB.101.060505}). Hofstadter-like systems, as prototypical examples of topological bands, have attracted considerable attention for their superconducting phases \cite{PhysRevLett.104.145301,Peotta2015,PhysRevB.104.184501,PhysRevB.108.035135,Shaffer2022,doi:10.1073/pnas.2426680122,chen2025,kuhlenkamp2025,zm39-dstj}, including both zero-momentum pairing and PDW states, due to the rich interplay between superconducting, quantum Hall, and spin liquid physics. While previous studies have investigated these systems\textemdash using mean-field theory, renormalization-group analysis, and numerical simulations\textemdash exactly solvable models exhibiting PDW ground states in topological bands have not yet been identified. More importantly, PDW states studied previously within mean-field or renormalization-group approaches are  typically weak-coupling PDWs, in the sense that they rely intrinsically on Fermi-surface nesting in the particle-particle (pp) channel. Here, by contrast, we construct exact strong-coupling PDW ground states in flat topological bands, where no 
Fermi surface exists. Our construction is fundamentally distinct from previous approaches: the pp-channel nesting responsible for the PDW order is “quantum geometric nesting” (QGN)~\cite{PhysRevX.14.041004}, arising purely from the structure of the Bloch wave functions.

In this Letter, we construct exactly solvable PDW ground states in Hofstadter-like systems with topological flat bands. Our construction utilizes a generalized QGN framework, an extension of a recently proposed concept that enables the systematic construction of exactly solvable ground states with intra-unit cell orders in flat band systems \cite{PhysRevX.14.041004}. 
We begin with a brief introduction to the generalized QGN used in our analysis. Then we consider a specific Hofstadter-like system possessing exact topological flat bands, the $\pi$-flux Kapit-Mueller model, and we construct exactly solvable interactions for which the PDW state is the exact ground state. After that, we prove that the generalized QGN in the PDW channel is protected by the non-commuting magnetic translation symmetry (MTS) and TRS, and hence our results hold for general Hofstadter-like systems with these two symmetries. Then we further generalize our results to Hofstadter-like systems with a general fractional flux $\phi=\frac{1}{p},p\geq 2$ (in units of $2\pi\hbar/e$) and finally discuss the possible experimental realizations of our models and implications for future studies.



{\bf Generalized QGN:} Consider a Hamiltonian $\hat{H}=\hat{H}_0+\hat{V}$, where  $\hat{H}_0$ hosts a set of flat bands forming a subspace $\mathcal{H}$.  When the electrons partially fill $\mathcal{H}$, then we can solve an effective Hamiltonian $\hat{H}_{\text{eff}}=\hat{P}\hat{V}\hat{P}$ instead of $\hat{H}$, where $\hat{P}$ is the projection to the subspace $\mathcal{H}$. This projection is exact in the limit $\frac{U}{\Delta}\rightarrow0$, where  $U,\Delta$ are the energy scales of the interaction $\hat{V}$ and  the band gap between the flat-band subspace $\mathcal{H}$ and other dispersive bands respectively. In practice, the method applies equally well to what we term ``relatively" flat bands satisfying $W\ll U\ll \Delta$, where $W$ is the bandwidth. In this regime, the ground state of  $\hat{H}_{\text{eff}}=\hat{P}\hat{V}\hat{P}$ remains an asymptotically exact ground state of $\hat{H}$ up to the order of $O(\frac{W}{U})$ and $O(\frac{U}{\Delta})$.

The (generalized) QGN provides an approach of constructing exactly solvable $\hat{H}_{\text{eff}}$ as follows. Here we mainly focus on superconducting ground states. We suppose the interaction is positive semi-definite $\hat{V}=\sum_{\mathbf{R},\mathbf{R}^{\prime},I,J}V_{IJ}(\mathbf{R}-\mathbf{R}^{\prime})(\hat{S}^{(I)}_{\mathbf{R}})^{\dagger}\hat{S}^{(J)}_{\mathbf{R}^{\prime}}$, where $\{\hat{S}_{\boldsymbol{R}}^{(I)}\}$ is a set of quadratic operators localized around the unit cell with the reference site $\mathbf{R}$, and the index $I=1,2...M$ labels the different possible choices of $\hat{S}_{\boldsymbol{R}}$ for each unit cell.  Written explicitly, $\hat{S}^{(I)}_{\mathbf{R}}=\sum_{\mu \nu, \boldsymbol{R}_{1} \boldsymbol{R}_{2}} S_{\mu \nu}^{(I)}\left(\boldsymbol{R}_{1}, \boldsymbol{R}_{2}\right) \hat{c}_{\boldsymbol{R}+\boldsymbol{R}_{1}, \mu}^{\dagger} \hat{c}_{\boldsymbol{R}+\boldsymbol{R}_{2}, \nu}$ with $\mu,\nu$ the sublattice indices (including orbital, spin and other internal degrees of freedom) and $S_{\mu \nu}^{(I)}\left(\boldsymbol{R}_{1}, \boldsymbol{R}_{2}\right) $ is a matrix in the sublattice and unit-cell space with a finite spatial range around $\mathbf{R}$  \footnote{In the original QGN framework, the operators 
$\{\hat{S}_{\boldsymbol{R}}^{(I)}\}$ are required to be Hermitian, whereas in the generalized QGN approach adopted here, we relax this Hermiticity condition.  }. If a superconducting order parameter $\hat{\eta}^{\dagger}=\sum_{\mathbf{k},\mu,\nu}\hat{c}_{\mu}^{\dagger}(\mathbf{k}+\frac{Q}{2})N_{\mu\nu}(\mathbf{k})\hat{c}^{\dagger}_{\nu}(-\mathbf{k}+\frac{\mathbf{Q}}{2})$ with the form factor $N_{\mu\nu}(\mathbf{k})$  satisfies $[\hat{P}\hat{\eta}^{\dagger}\hat{P},\hat{P}\hat{S}^{(I)}_{\mathbf{R}}\hat{P}]=0$ for any $\hat{S}^{(I)}_{\mathbf{R}}$, then the superconducting ground state of the generalized $\eta$-pairing form: $|\psi\rangle=(\hat{P}\hat{\eta}^{\dagger}\hat{P})^N|\text{vac}\rangle$  is an exact ground state of $\hat{H}^{\prime}=\sum_{\mathbf{R},\mathbf{R}^{\prime},I,J}V_{IJ}(\mathbf{R}-\mathbf{R}^{\prime})(\tilde{S}^{(I)}_{\mathbf{R}})^{\dagger}\tilde{S}^{(J)}_{\mathbf{R}^{\prime}}$. Here $\hat{P}\hat{S}^{(I)}_{\mathbf{R}}\hat{P}=\tilde{S}^{(I)}_{\mathbf{R}}+\langle\hat{S}^{(I)}_{\mathbf{R}}\rangle$, where $\tilde{S}^{(I)}_{\mathbf{R}}$ contains only fermion operators projected onto the flat bands, and $\langle\hat{S}^{(I)}_{\mathbf{R}}\rangle$ is the expectation value evaluated on the fully filled lower dispersive bands, which vanishes when the flat bands are the lowest bands.  $\hat{H}^{\prime}$ differs from 
the effective Hamiltonian $\hat{H}_{\text{eff}}=\hat{P}\hat{V}\hat{P}$ by a quadratic term since the interaction $\hat{V}$ is not normal-ordered.  A sufficient condition for $|\psi\rangle$ to be an exact ground state of $\hat{H}_{\text{eff}}$ is that the arising quadratic term is a trivial chemical potential term and can be omitted in a particle-number conserved system. This condition, however, is not automatically satisfied and must be checked explicitly for each concrete model. Here we want to remark that in the original QGN framework \cite{han2025}, the order parameter $\hat{\eta}^{\dagger}$ is restricted to be intra-unit-cell, so that the associated form factor $N_{\mu\nu}$ is independent of momentum. In this work, we generalize the framework to order parameters with momentum-dependent form factors $N_{\mu\nu}(\mathbf{k})$, thereby allowing us to construct exactly solvable PDW states in topological bands. 

Although $[\hat{P}\hat{\eta}^{\dagger}\hat{P},\hat{P}\hat{S}^{(I)}_{\mathbf{R}}\hat{P}]=0,\forall \mathbf{R},I$ are generally difficult to satisfy for generic bands, a simple sufficient criterion for it is  $[\hat{\eta}^{\dagger},\hat{S}^{(I)}_{\mathbf{R}}]=0$, provided that $\hat{\eta}^{\dagger}$ satisfies the following generalized QGN condition:
\begin{equation} 
\begin{split} 
\hat{\eta}^{\dagger} = \sum_{\mathbf{k},m,n} \hat{\gamma}_{m}^{\dagger}\!\left(\mathbf{k}+\frac{\mathbf{Q}}{2}\right) N_{mn}(\mathbf{k}) \hat{\gamma}_{n}^{\dagger}\!\left(-\mathbf{k}+\frac{\mathbf{Q}}{2}\right), \\ 
N_{mn}(\mathbf{k}) =0 \quad \text{unless } \bigl(m,n\in\mathrm{flat}) \text{ or } \bigl(m, n \notin\mathrm{flat}\bigr).
\end{split} 
\label{eq:qgn}
\end{equation}
where  $\hat{\gamma}_{m}^{\dagger}(\mathbf{k})$ are fermion operators acting in the band $m$.  The proof is straightforward, since $[\hat{\eta}^{\dagger},\hat{S}^{(I)}_{\mathbf{R}}]=0$, we have $\hat{P} [\hat{\eta}^{\dagger},\hat{S}^{(I)}_{\mathbf{R}}]\hat{P}=[\hat{P}\hat{\eta}^{\dagger}\hat{P},\hat{P}\hat{S}^{(I)}_{\mathbf{R}}\hat{P}]=0$, where we use the fact that  $\hat{P}\hat{\eta}^{\dagger}(1-\hat{P})\hat{S}^{(I)}_{\mathbf{R}}\hat{P}=\hat{P}\hat{S}^{(I)}_{\mathbf{R}}(1-\hat{P})\hat{\eta}^\dagger\hat{P}=0$. 
These two identities follow  because $(1-\hat{P})\hat{S}^{(I)}_{\mathbf{R}}\hat{P},\hat{P}\hat{S}^{(I)}_{\mathbf{R}}(1-\hat{P})$ select only those intermediate states in $1-\hat{P}$ with exactly one electron or hole in the dispersive bands, and the generalized QGN condition in Eq.~\eqref{eq:qgn} then forces the two operators $\hat{P}\hat{\eta}^{\dagger}(1-\hat{P})\hat{S}^{(I)}_{\mathbf{R}}\hat{P},\hat{P}\hat{S}^{(I)}_{\mathbf{R}}(1-\hat{P})\hat{\eta}^\dagger\hat{P}$ to vanish.  In practice, requiring $[\hat{\eta}^{\dagger},\hat{S}^{(I)}_{\mathbf{R}}]=0$ simplifies the construction of $\hat{S}^{(I)}_{\mathbf{R}}$.


Physically, the generalized QGN describes a ``nesting" relation of two flat-band wave functions at momenta $\pm\mathbf{k}+\frac{\mathbf{Q}}{2}$, since they are connected by the matrix $N_{\mu\nu}(\mathbf{k})$.  Following \cite{PhysRevX.14.041004} we refer to this as “quantum geometric nesting” because it does not rely on the existence of a Fermi surface or on details of the band dispersion; rather, it is determined solely by  overlaps of the wave functions. 



Here, we construct exactly solvable PDW states in topological flat bands within the framework of the generalized QGN, and demonstrate that a PDW order parameter satisfying the generalized QGN condition can generally be constructed when the system possesses  MTS. 

{\bf Model:} We consider a spin-$\frac{1}{2}$ fermionic system described by $\hat{H}=\hat{H}_0+\hat{V}$, where 
the (noninteracting) 
$\hat{H}_0$ spectrum contains one (relatively) flat band per spin  with a nonzero Chern number. Furthermore, we impose that the full Hamiltonian $\hat{H}$ preserves MTS and TRS, which enables us to construct an (asymptotically) exactly solvable interaction $\hat{V}$. 

A prototypical example of the desired flat-band Bloch states is provided by the Landau-level wavefunctions, which can serve as exact zero modes of a positive semi-definite parent Hamiltonian, such as the Kapit–Mueller (KM) model and its generalizations~\cite{PhysRevLett.105.215303,shen2025}. Consequently, we begin with the time-reversal-symmetric version of the KM model \cite{PhysRevLett.105.215303}:  $\hat{H}_0=\sum_{\mathbf{r}_i,\mathbf{r}_j\in\Lambda,\sigma}J_{\sigma}(\mathbf{r}_i,\mathbf{r}_j)\hat{c}^{\dagger}_{\mathbf{r}_i,\sigma}\hat{c}_{\mathbf{r}_j,\sigma}$, where $\Lambda$ represents the points on a square lattice $\mathbf{r}_i=x_i\mathbf{a}_x+y_i\mathbf{a}_y$ with $\mathbf{a}_x,\mathbf{a}_y$ are lattice unit vectors and without loss of generality, we set the lattice constants $a_x=a_y=1$ hereafter. We use $\sigma=\pm $ and $\uparrow,\downarrow$ interchangeably to denote spin up and down. The hopping amplitudes $J_{\uparrow}(\mathbf{r}_i,\mathbf{r}_j)=J^*_{\downarrow}(\mathbf{r}_i,\mathbf{r}_j)=W(\mathbf{r}) e^{-i\phi\pi(y_i+y_j)x}$  decay exponentially with the relative distance $|\mathbf{r}|=|\mathbf{r}_i-\mathbf{r}_j|$: $ W(\mathbf{r})=t(-1)^{x+y+xy}e^{-\frac{\pi}{2}(1-\phi)r^2}$,
where we take the Landau gauge and $\phi=\frac{q}{p}<1$  (with $p,q \in \mathbb{Z}^+$) is the magnetic flux (in the unit of $\frac{2\pi\hbar}{e}$) per plaquette.  We  start from the simplest $\pi$-flux ($\phi=\frac{1}{2}$) case, and each magnetic unit cell contains two sites. Since $\hat{H}_0$ commutes with the lattice translation symmetries $\hat{T}_{a_x},\hat{T}_{2a_y}$, we take the magnetic unit cell as: $\mathbf{r}_A=\mathbf{R}=x\mathbf{a}_x+2y\mathbf{a}_y,\,\,\mathbf{r}_B=\mathbf{r}_A+\mathbf{a}_y,\,\, x,y\in \mathbb{Z}$, where we label the two sites as $A$, $B$. The positive semi-definite $\hat{H}_0$ has an exact flat band (per spin) with zero energy whose  Bloch states are lowest LL wavefunctions  sampled on the lattice $\Lambda$ with periodic boundary conditions  \cite{PhysRevLett.105.215303}.

As a result, we  denote the normalized flat-band Bloch states as: 
\begin{equation}
\begin{aligned}
|u(\mathbf{k}),\uparrow\rangle&=a(\mathbf{k})|\mathbf{k},A,\uparrow\rangle+b(\mathbf{k})|\mathbf{k},B,\uparrow\rangle,\\
     |u(\mathbf{k}),\downarrow\rangle&=a^*(-\mathbf{k})|\mathbf{k},A,\downarrow\rangle+b^*(-\mathbf{k})|\mathbf{k},B,\downarrow\rangle,
\end{aligned}
\end{equation}
 The sublattice bases are: $|\mathbf{k},A,\sigma\rangle=\frac{1}{\sqrt{S}}\sum_{\mathbf{R}}e^{i\mathbf{k}\cdot\mathbf{R}}|\mathbf{R},\sigma\rangle,|\mathbf{k},B,\sigma\rangle=\frac{1}{\sqrt{S}}\sum_{\mathbf{R}}e^{i\mathbf{k}\cdot(\mathbf{R}+\mathbf{a}_y)}|\mathbf{R}+\mathbf{a}_y,\sigma\rangle$, where the Wannier states are normalized $\langle\mathbf{r},\sigma|\mathbf{r}^{\prime},\sigma^{\prime}\rangle=\delta_{\mathbf{r},\mathbf{r}^{\prime}}\delta_{\sigma,\sigma^{\prime}}$. The momentum $k_{x}=\frac{2\pi m_{x}}{N_x},k_{y}=\frac{\pi m_{y}}{N_y},m_{x/y}=1,2...N_{x/y}$ and $N_{x/y}$ are the number of unit cells in the $x$ and $y$-directions respectively. $S=N_xN_y$ is the total number of magnetic unit cells. 
The flat bands are topologically nontrivial with nonzero Chern number $C_{\uparrow}=1,C_{\downarrow}=-1$. Moreover, the flat band  is an ideal  Kähler  band satisfying the trace bound \cite{shen2025,PhysRevLett.127.246403,1zg9-qbd6}.

{\bf Exactly solvable PDW:}  
Using the generalized QGN framework, we first construct the PDW order parameter satisfying the generalized QGN condition. Our construction is intrinsically based on the magnetic translation symmetries, whose generators are: $\hat{\tilde{T}}_{a_x}=\hat{T}_{a_x},\hat{\tilde{T}}_{a_y}=\sum_{\mathbf{r}}(-1)^{r_x}|\mathbf{r}+\mathbf{a}_y\rangle\langle\mathbf{r}|$, such that  $\hat{\tilde T}_{a_y}^2 = \hat T_{2 a_y}$. Since these two generators anti-commute $\{\hat{\tilde{T}}_{a_x},\hat{\tilde{T}}_{a_y}\}=0$, the momentum of the Bloch states $| u(\mathbf{k}),\sigma\rangle$ will be shifted under the application of $\hat{\tilde{T}}_{-a_y}$: $\hat{\tilde{T}}_{-a_y}|u(\mathbf{k}),\sigma\rangle\propto |u(\mathbf{k}+\mathbf{Q}),\sigma\rangle$, where $\mathbf{Q}=(\pi,0)$. Written explicitly, we have:
\begin{equation}
\begin{array}{l}
a(\mathbf{k})=e^{i\theta_{\mathbf{k}}}e^{-ik_ya_y}b(\mathbf{k}+\mathbf{Q}) \\ b(\mathbf{k})=e^{i\theta_{\mathbf{k}}}e^{-ik_ya_y}a(\mathbf{k}+\mathbf{Q})  \end{array}
    \label{Bloch}
\end{equation}
where $\theta_{\mathbf{k}}$ are model-dependent parameters and we verify in Section A of the supplementary materials that the phase $\theta_{\mathbf{k}}=0$ in the  KM model considered here \cite{SupMat}. This is a manifestation of “nesting” of Bloch wavefunctions even in the absence of a Fermi surface: two flat-band wavefunctions separated by momentum $\mathbf{Q}$ are connected by a unitary matrix, which is a purely quantum geometric effect guaranteed by the MTS. 

The wave function property in Eq. \eqref{Bloch} strongly suggests an $S_z=0$ inter-sublattice pairing with pairing momentum $\mathbf{Q}$ could satisfy the generalized QGN, or be block-diagonalized in the band basis, since the center-of-mass momentum $\mathbf{Q}$ is compensated by sublattice exchange, and then the two Bloch wavefunctions of the order parameter share the same momentum. Indeed, the following PDW order parameter satisfies the generalized QGN \cite{SupMat}:
\begin{equation}
\eta^{\dagger}\!=\!\!\sum_{\mathbf{R},\sigma,\sigma^{\prime}}\!\!(-1)^{R_x}\!\left[\hat{c}_{\sigma}^{\dagger}(\mathbf{R}+\mathbf{a}_y)-\hat{c}_{\sigma}^{\dagger}(\mathbf{R}-\mathbf{a}_y)\right]\!(\sigma_x)_{\sigma,\sigma^{\prime}}\hat{c}_{\sigma^{\prime}}^{\dagger}(\mathbf{R}),
    \label{pdw}
\end{equation}
where $\sigma_x$ acts on the spin index.

Further, we can also construct the following set of $\hat{S}_{\mathbf{R}}$ which commutes with $\eta^{\dagger}$:
\begin{equation}
\begin{aligned}
     \hat{S}^{(1)}_{\mathbf{R}}&= \mbox{$\sum_{\sigma}$} \left[\hat{c}_{\sigma}^{\dagger}(\mathbf{R}+\mathbf{a}_y)-\hat{c}_{\sigma}^{\dagger}(\mathbf{R}-\mathbf{a}_y)\right]\hat{c}_{\sigma}(\mathbf{R}),\\
      \hat{S}^{(2)}_{\mathbf{R}}&=\mbox{$\sum_{\sigma}$}\left[\hat{c}_{\sigma}^{\dagger}(\mathbf{R}+2\mathbf{a}_y)-\hat{c}_{\sigma}^{\dagger}(\mathbf{R})\right]\hat{c}_{\sigma}(\mathbf{R}+\mathbf{a}_y),
\end{aligned} 
\label{block}
\end{equation}
which also guarantee $\left[\hat{P}\hat{S}^{(1,2)}_{\mathbf{R}}\hat{P},\hat{P}\eta^{\dagger}\hat{P}\right]=0$ as $\eta^{\dagger}$ satisfies the generalized QGN. As a result, our constructed interaction is:
\begin{equation}
\begin{aligned}
        \hat{V}&=U\sum_{\mathbf{R};I=1,2}(\hat{S}^{(I)}_{\mathbf{R}})^{\dagger}\hat{S}^{(I)}_{\mathbf{R}}\\
        &=-U\sum_{\mathbf{r}\in\Lambda}\left(\hat{n}(\mathbf{r})\hat{n}_{\text{anti}}(\mathbf{r})+\vec{\hat{S}}(\mathbf{r})\cdot \vec{\hat{S}}_{\text{anti}}(\mathbf{r})\right)+2U\hat{N},
\end{aligned}
\label{eq:int}
\end{equation}
where $U>0$ is the interaction strength  and $2U\hat{N}$ is a constant in systems with fixed electron number which can be neglected. $\hat{n}_{\text{anti}}(\mathbf{r})$ is the density of the anti-bonding orbital formed by the two nearest neighbours of site $\mathbf{r}$: $\hat{n}_{\text{anti}}(\mathbf{r})=\sum_{\sigma}\hat{c}^{\dagger}_{\text{anti},\sigma}(\mathbf{r})\hat{c}_{\text{anti},\sigma}(\mathbf{r}),\hat{c}_{\text{anti},\sigma}(\mathbf{r})=\frac{1}{\sqrt{2}}\left(\hat{c}_{\sigma}(\mathbf{r}+\mathbf{a}_y)-\hat{c}_{\sigma}(\mathbf{r}-\mathbf{a}_y)\right)$. $\vec{\hat{S}}(\mathbf{r}),\vec{\hat{S}}_{\text{anti}}(\mathbf{r})$ are the spin operators: $\vec{\hat{S}}(\mathbf{r})=\hat{c}^{\dagger}(\mathbf{r})\vec{\sigma}\hat{c}(\mathbf{r}),\vec{\hat{S}}_{\text{anti}}(\mathbf{r})=\hat{c}_{\text{anti}}^{\dagger}(\mathbf{r})\vec{\sigma}\hat{c}_{\text{anti}}(\mathbf{r})$. The effective Hamiltonian $\hat{H}_{\text{eff}}=\hat{P}\hat{V}\hat{P}$ yields the exact PDW ground state in the form of a generalized $\eta$-pairing state: $|\psi\rangle=(\hat{P}\eta^{\dagger}\hat{P})^{N}|\text{vac}\rangle$, where the system has $2N$ electrons in total and we show this state has off-diagonal long-range order with electron density $0<\frac{2N}{S}<2$ in the thermodynamic limit in the supplementary material \cite{SupMat}. Written explicitly, the projected PDW order parameter takes the form:
\begin{equation}
\begin{aligned}
       \hat{P}\eta^{\dagger}\hat{P}&=\sum_{\mathbf{k},\sigma,\sigma^{\prime}}\hat{\gamma}_{\sigma}^{\dagger}(\mathbf{k}+\frac{\mathbf{Q}}{2})F_{\sigma,\sigma^{\prime}}(\mathbf{k})(\hat{\gamma}_{\sigma^{\prime}}^{\dagger}(-\mathbf{k}+\frac{\mathbf{Q}}{2}))^T,\\
       F_{\sigma,\sigma^{\prime}}(\mathbf{k})&=-2i(\sigma_x)_{\sigma,\sigma^{\prime}}\sin(k_y)e^{-i\sigma k_y},
\end{aligned}
\label{eq:eta_pairing_form_factor}
\end{equation}
where the two-component operator $\hat{\gamma}_{\sigma}^{\dagger}(\mathbf{k})$ is the creation operator in the flat band basis. As we can see from the form factor $F(\mathbf{k})$, the PDW is a mixed spin-singlet and triplet pairing, since the spin rotation symmetry of the total Hamiltonian $\hat{H}$ is $U(1)_z$ and only the total $S_z$ is a good quantum number.


An additional issue concerns the quadratic term difference between $\hat{H}_{\text{eff}}=\hat{P}\hat{V}\hat{P}$ and $U\sum_{\mathbf{R};I=1,2}(\hat{P}\hat{S}^{(I)}_{\mathbf{R}}\hat{P})^{\dagger}\left(\hat{P}\hat{S}^{(I)}_{\mathbf{R}}\hat{P}\right)$. We prove in the supplementary material that the quadratic terms here are in the form of a chemical potential in the flat band space \cite{SupMat}, and hence can be omitted in a particle-number conserved system.

{\bf Universality and uniqueness of the exactly solvable PDW:} Our construction of the exactly solvable PDW is not a special feature at the  KM point, but is universal in $\pi$-flux systems with the same MTS and time-reversal symmetries, e.g. the generalized KM model whose flat band wavefunctions are $n$-th Landau level wavefunctions \cite{shen2025}. Since in such $\pi$-flux systems, the wave functions still satisfy Eq.~\eqref{Bloch} with a nonzero $\theta_{\mathbf{k}}$ in general, and the (unprojected) PDW order parameter $\eta^{\dagger}$ in Eq.~\eqref{pdw} still satisfies the generalized QGN \cite{SupMat}. Besides,  for any such $\hat{H}_0$ possessing MTS and TRS, with (relatively) flat bands as the lowest bands, the quadratic term arising from projection of the above interaction $\hat{V}$, which is comprised of $\hat{S}_{\mathbf{R}}^{(I)}$ operators in Eq.~\eqref{block} is always a trivial chemical potential term;  a proof is provided in the Supplementary Material \cite{SupMat}. As a result,  the PDW state $|\psi\rangle=(\hat{P}\eta^{\dagger}\hat{P})^N|\text{vac}\rangle$ is still  an exact ground state of the effective Hamiltonian $\hat{H}_{\text{eff}}=\hat{P}\hat{V}\hat{P}$, where $\eta^{\dagger}$ is given by Eq. \eqref{pdw}, and the flat-band projection $\hat{P}$ now depends on our choice of the kinetic energy $\hat{H}_0$. 

As a result, we are free to truncate and tune the hopping amplitudes of the  KM model and use the same interaction $\hat{V}$, and then our constructed Hamiltonian $\hat{H}=\hat{H}_0+\hat{V}$ still admits an asymptotically exactly solvable PDW ground state $|\psi\rangle=(\hat{P}\eta^{\dagger}\hat{P})^N|\text{vac}\rangle$, as long as we still have a relatively flat band ($W\ll U\ll \Delta$) per spin.

\begin{figure}
    \centering
    \includegraphics[width=0.75\linewidth]{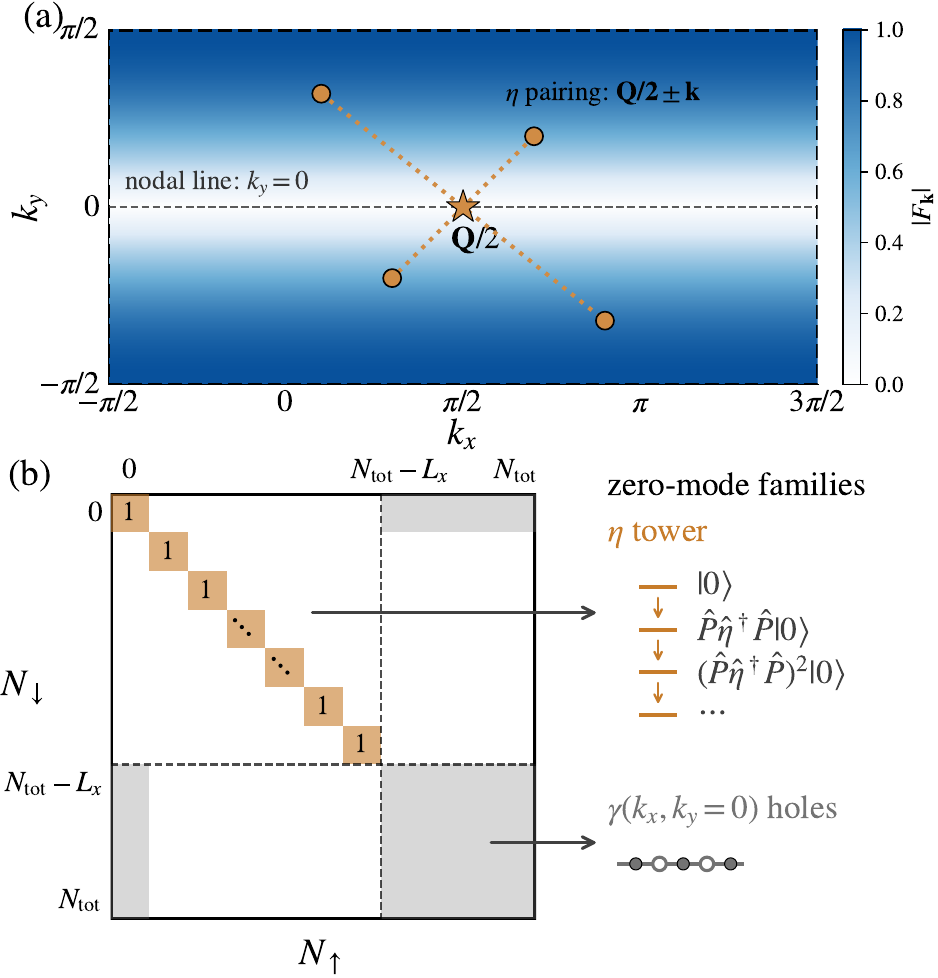}
    \caption{Schematic illustration of zero modes in the system.
(a) Magnitude of the $\eta$-pairing structure factor
$|F_{\mathbf{k}}|=\sin(k_y)e^{ik_y}$, which vanishes along the nodal line
$k_y=0$. Orange markers denote paired momenta  with center-of-mass momentum $\mathbf{Q}$.
(b) Zero-mode families in the filling sectors $(N_\uparrow,N_\downarrow)$.
In the accessible system sizes, the ED results are consistent with the two types of zero-modes analytically identified \cite{SupMat}.
$N_{\rm{tot}}=L_xL_y/2$ denote the total number of momenta of MBZ.
For $N_{\uparrow,\downarrow}<N_{\rm{tot}}-L_x$, the identified zero modes occur only
on the diagonal $N_\uparrow=N_\downarrow$ and correspond to the
$\eta$-pairing states. The shaded gray regions denote hole-like zero modes
on the nodal line, corresponding to $\hat{\gamma}_{\sigma}(k_x,k_y=0)$ acting on a fully filled band.  In the diagonal shaded region where
both spin species support hole-like zero modes ($N_{\uparrow,\downarrow}\geq N_{\rm{tot}}-L_x$), the
degeneracy is
$\binom{L_x}{N_{\rm{tot}}-N_\uparrow}\binom{L_x}{N_{\rm{tot}}-N_\downarrow}$, while in the two
remaining shaded regions ($N_{\uparrow/\downarrow}\geq N_{\rm{tot}}-L_x,N_{\downarrow/\uparrow}=0$), the degeneracy is
$\binom{L_x}{N_{\rm{tot}}-N_{\uparrow/\downarrow}}$.
}
    \label{fig:Fig1}
\end{figure}

The remaining question is whether the exact PDW state is the unique ground state of our model. We address this by numerically computing the ground state degeneracy (GSD) of $\hat{H}_{\text{eff}}=\hat{P}\hat{V}\hat{P}$, with $\hat{P}$ projecting to the flat-band space of the KM model, using exact diagonalization (ED). The PDW state is indeed the unique ground state at generic fillings, except when the band is nearly fully occupied~\cite{SupMat}, as is illustrated in Fig.~\ref{fig:Fig1}. The appearance of additional ground states near full filling can be understood from the structure of $\hat{P}\hat{S}^{(I)}_{\mathbf{R}}\hat{P}=\sum_{\mathbf{p},\mathbf{q}}e^{i\mathbf{R}(\mathbf{q}-\mathbf{p})}\hat{\gamma}^{\dagger}(\mathbf{p})S^{(I)}(\mathbf{p},\mathbf{q})\hat{\gamma}(\mathbf{q})$, whose form factor $S^{(I)}(\mathbf{p},\mathbf{q})$ contains a $\sin(p_y)$ factor. Consequently, if all the states away from the $k_x$ axis are fully occupied, then any state with a partially filled $k_x$-axis is annihilated by $\hat{P}\hat{S}^{(I)}_{\mathbf{R}}\hat{P},\forall\mathbf{R},I$, and is therefore a ground state of $\hat{H}_{\text{eff}}=\hat{P}\hat{V}\hat{P}$. To verify this interpretation, we add an energy penalty term $\mu\sum_{k_y=0;\sigma}\hat{\gamma}_{\sigma}^{\dagger}(\mathbf{k})\hat{\gamma}_{\sigma}(\mathbf{k}),\mu>0$ to $\hat{H}_{\text{eff}}$, and these additional ground states are lifted~\cite{SupMat}. More importantly, these extra ground states occur only within the measure-zero filling intervals $\nu\in[1-\frac{1}{L_y},1]\cup[2-\frac{2}{L_y},2]$ in the thermodynamic limit. Thus, the PDW state is the unique ground state at generic fillings. The appearance of multiple zero modes for a nearly filled band does not imply a large number of `near zero modes' in the excitation spectrum of the PDW state. As we illustrated the density of states obtained from ED calculations~\cite{SupMat}, the low energy spectrum is relatively sparse and the peak of the density of states is approximately located at the middle of the whole spectrum in the PDW regime. 

{\bf Generalizations to other fluxes:} Our results can be naturally generalized to Hofstadter systems with fractional flux $\phi=\frac{1}{p},p\geq 2$. We still require the system to preserve MTS and time-reversal symmetries, and there is a  single relatively flat band per spin. A typical example is the KM model, and the generators of MTS take $\hat{\tilde{T}}_{a_x}=\hat{T}_{a_x},\hat{\tilde{T}}_{a_y}=\sum_{\mathbf{r}\in\Lambda}e^{-i\sigma\frac{2\mathbf{Q}\cdot\mathbf{r}}{p}}|\mathbf{r}+\mathbf{a}_y,\sigma\rangle\langle\mathbf{r},\sigma|$ with $\mathbf{Q}=(\pi,0)$. A PDW order parameter with center-of-mass momentum $\frac{2\mathbf{Q}}{p}$ satisfying the generalized QGN can be constructed naturally by conjugating one fermion operator by the MTS $\hat{\tilde{T}}_{a_y}$ as: $\eta_p^{\dagger}=\sum_{\mathbf{r}}\hat{\tilde{c}}_{\downarrow}^{\dagger}(\mathbf{r})\hat{c}_{\uparrow}^{\dagger}(\mathbf{r})=\sum_{\mathbf{r}}e^{i\frac{2\mathbf{Q}\cdot\mathbf{r}}{p}}\hat{c}_{\downarrow}^{\dagger}(\mathbf{r}+\mathbf{a}_y)\hat{c}_{\uparrow}^{\dagger}(\mathbf{r}),$ where $\hat{\tilde{c}}_{\downarrow}^{\dagger}(\mathbf{r})=\hat{\tilde{T}}_{a_y}\hat{c}_{\downarrow}^{\dagger}(\mathbf{r})\hat{\tilde{T}}^{-1}_{a_y}=\hat{c}_{\downarrow}^{\dagger}(\mathbf{r}+\mathbf{a}_y)e^{i\frac{2\mathbf{Q}\cdot\mathbf{r}}{p}}$. We verify this PDW order parameter satisfies the generalized QGN in the supplementary material \cite{SupMat}.

The exactly solvable interaction $\hat{V}$ can also be constructed from a set of $\{\hat{S}^{(\mu)}_{\mathbf{R}}\}$ operators which commute with $\eta_p^{\dagger}$ (both before and after the projection to the flat band):
\begin{equation}
 \label{eq:int_p}
\hat{V}=U\sum_{\mathbf{R}}\sum_{\mu=0}^{n-1}\left((S^{(\mu)}_{\mathbf{R}})^{\dagger}S^{(\mu)}_{\mathbf{R}}+S^{(\mu)}_{\mathbf{R}}(S^{(\mu)}_{\mathbf{R}})^{\dagger}\right),\\
\end{equation}
$$
S^{(\mu)}_{\mathbf{R}}=\sum_\sigma \sigma \hat{c}_{
\sigma}^{\dagger}(\mathbf{R}+\mu\mathbf{a}_y)\hat{c}_{\sigma}(\mathbf{R}+(\mu-\sigma)\mathbf{a}_y)
$$
where $\sigma=1,-1$ should be interpreted as $\uparrow,\downarrow$ and 
$\mu=0,1,\ldots, p-1$  labels the sublattice. The set of sites $\mathbf{R}=x\mathbf{a}_x+py\mathbf{a}_y,\quad x,y\in \mathbb{Z},$ represents the lattice of magnetic unit cells and encodes the corresponding lattice translation symmetries. The effective Hamiltonian $\hat{H}_{\text{eff}}=\hat{P}\hat{V}\hat{P}$ has the following exact PDW ground state in the generalized $\eta$-pairing form: $|\psi\rangle=(\hat{P}\eta_p^{\dagger}\hat{P})^N|\text{vac}\rangle$ \cite{SupMat}, where $\hat{P}$ is the projection to the flat band space of the kinetic energy. Similar to the $\pi$-flux systems, the exactly solvable PDW state presented here is not a special feature of the KM model but generally arises in systems possessing the same MTS and TRS with the flat bands as the lowest bands \cite{SupMat}. 

 The interaction in Eq.~\eqref{eq:int_p} is different from that in Eq.~\eqref{eq:int} when $p=2$, since they have different ground states, as reflected by the distinct order parameters  $\hat{\eta}^{\dagger}_{p}$ and $\hat{\eta}^{\dagger}$. Moreover, only $\hat{\eta}^{\dagger}_{p}$ admits a natural generalization to $p>2$, since they satisfy the generalized QGN for any $p\geq 2$.

{\bf Discussions and concluding remarks:} We construct exactly solvable PDW ground states in topological flat bands using the generalized QGN framework. Our exactly solvable model provides a valid platform for further studies of the PDW physics in topological bands, such as the excitation spectrum, superfluid stiffness and exotic transport phenomena of PDW state, such  as the Josephson effect \cite{annurev}, etc.  Promising experimental platforms for realizing our models include ultracold atom systems \cite{PhysRevLett.111.185301}, where time-reversal-symmetric Hofstadter Hamiltonians can be engineered. Another viable candidate is twisted homobilayer transition metal dichalcogenides (TMDs) \cite{PhysRevLett.122.086402,PhysRevLett.132.096602}, in which an effective magnetic field is induced by an applied gate potential, thereby preserving time-reversal symmetry. When the periodic gate potential forms a meron lattice, the resulting moiré material effectively realizes the $\pi$-flux configuration required by our construction. 


Finally, we comment that we can ease the requirement of time-reversal symmetry and only require the non-commuting MTS to construct exactly solvable PDW states. Adopting the ansatz in \cite{han2025}, we can also construct an exactly solvable minimal model of spinless fermions whose ground state is a topological FFLO phase. This phase has a nonzero Chern number and, consequently, supports a chiral Majorana edge mode \cite{SupMat}.

{\bf Acknowledgments:} We sincerely thank Steven Kivelson, Sid Parameswaran, Ziwei Wang, Hong Yao, Jia-Xin Zhang for helpful discussions, and especially Zhaoyu Han for many valuable discussions. This work is supported by the EPSRC under grant EP/X030881/1. The Flatiron Institute is a division of the Simons Foundation. 

\bibliography{bib}
\clearpage
\newpage
\widetext

\title {Supplementary material for `Exactly solvable pair-density wave  in topological bands from magnetic translation symmetries'}
\date{\today}

\maketitle

\onecolumngrid

\setcounter{equation}{0}
\setcounter{figure}{0}
\setcounter{table}{0}
\makeatletter
\renewcommand{\theequation}{S\arabic{equation}}
\renewcommand{\thefigure}{S\arabic{figure}}
\renewcommand{\bibnumfmt}[1]{[S#1]}
\renewcommand{\citenumfont}[1]{S#1}
 \section{A. Proof of the generalized QGN, off-diagonal long-range order and the triviality of the quadratic term}
 We begin with the $\pi$-flux systems and prove that our constructed PDW order parameter $\hat{\eta}^{\dagger}$ satisfies the generalized QGN, the PDW ground state has ODLRO, and the quadratic term arising from the projection of our constructed interaction $\hat{V}$ is a chemical potential term which can be neglected. 

\subsection{ $A_1$. $\pi$-flux systems}
Our conclusions in this section apply generally to spin-$\frac{1}{2}$ $\pi$-flux systems with time-reversal symmetry and (relatively) flat bands. We begin with a representative class of models, namely the generalized KM model whose exact flat-band wavefunctions (also exact ground states) are the n-th Landau level wavefunctions \cite{shen2025}, and then show that the properties derived below rely only on the presence of MTS and time-reversal symmetry.
\subsubsection{Generalized  QGN}

First, our conventions for the model and operators are as follows. There are two sublattices in real space which we denote as $\mu=A,B$ and there are two bands per spin, and we denote the normalized flat band wave functions as: 
\begin{equation}
\begin{aligned}
     |u(\mathbf{k}),\uparrow\rangle&=a_n(\mathbf{k})|\mathbf{k},A,\uparrow\rangle+b_n(\mathbf{k})|\mathbf{k},B,\uparrow\rangle,\\
     |u(\mathbf{k}),\downarrow\rangle&=a_n^*(-\mathbf{k})|\mathbf{k},A,\downarrow\rangle+b_n^*(-\mathbf{k})|\mathbf{k},B,\downarrow\rangle,
\end{aligned}
\end{equation}
We adopt the following convention of Fourier transformation: $\hat{c}^{\dagger}({\mathbf{k}})=\frac{1}{\sqrt{S}}\sum_{\mathbf{r}}\hat{c}^{\dagger}(\mathbf{r})e^{i\mathbf{k}\cdot\mathbf{r}}$, where $S$ is the total number of magnetic unit cells. We denote the fermion operators in the energy band basis as $\hat{\gamma}(\mathbf{k})$, and the unitary transformation between $\hat{\gamma}(\mathbf{k})$ and $\hat{c}(\mathbf{k
})$ is: $\hat{\gamma}^{\dagger}(\mathbf{k})=\hat{c}^{\dagger}(\mathbf{k})U(k)$, with the matrix $U(k)$:
\begin{equation}
   U^{\uparrow}(\vec{k})=(U^{\downarrow}(-\vec{k}))^*= \begin{pmatrix}
         a_n(\vec{k}) & -b_n^{*}(\vec{k}) \\
        b_n(\vec{k}) & a_n^*(\vec{k})
    \end{pmatrix},
\end{equation}
where $U^{\sigma}_{\mu1},U^{\sigma}_{\mu2}$ are the Bloch functions of the flat band and the dispersive band respectively. 

Due to the anti-commutating MTS $\{\hat{\tilde{T}}_{a_x},\hat{\tilde{T}}_{a_y}\}=0$ , the Bloch functions satisfy:
\begin{equation}
    a_n(\mathbf{k})=e^{i\theta_{\mathbf{k}}}e^{-ik_ya_y}b_n(\mathbf{k}+\mathbf{Q}),b_n(\mathbf{k})=e^{i\theta_{\mathbf{k}}}e^{-ik_ya_y}a_n(\mathbf{k}+\mathbf{Q}),
\end{equation}
where $\mathbf{Q}=(\pi,0)$. It can be directly verified that  the phase factors $\theta_{\mathbf{k}}$ are zero if $\langle\mathbf{r}|u(\mathbf{k})\rangle$ are (normalized) n-th Landau level wavefunctions sampled on the lattice:
\begin{equation}
    \begin{aligned}
     a_{n}(\vec{k})&=\frac{\psi_{n,\mathbf{k}}(0)}{\sqrt{|\psi_{n,\mathbf{k}}(0)|^2+|\psi_{n,\mathbf{k}}(y=a_y)|^2}},\quad b_{n}(\vec{k})=e^{-ik_ya_y}\frac{\psi_{n,\mathbf{k}}(y=a_y)}{\sqrt{|\psi_{n,\mathbf{k}}(0)|^2+|\psi_{n,\mathbf{k}}(y=a_y)|^2}}\\
       \psi_{n,\mathbf{k}}(0)&=  \sum_{m \in \mathbb{Z}} H_{n}\left(k_{x} \ell_B+\left(2 \pi \ell_B / a_{x}\right) m\right) e^{-(1 / 2)\left(k_{x} \ell_B+\left(2 \pi \ell_B / a_{x}\right) m\right)^{2}} e^{-2i k_{y} a_{y} m},\\
       \psi_{n,\mathbf{k}}(y=a_y)&=\sum_{m \in \mathbb{Z}} H_{n}\left(a_y/l_B+k_{x} \ell_B+\left(2 \pi \ell_B / a_{x}\right) m\right) e^{-(1 / 2)\left(a_y/l_B+k_{x} \ell_B+\left(2 \pi \ell_B / a_{x}\right) m\right)^{2}} e^{-2i k_{y} a_{y} m},
    \end{aligned}
\end{equation}
where the magnetic length $l^2_B=\frac{a_xa_y}{\pi}$. They are the exact ground states of the following generalized KM model \cite{shen2025}:
\begin{equation}
\begin{aligned}
     &\hat{H}_{0,n}=\sum_{\sigma}\hat{D}_{n,\sigma}^{\dagger}\hat{D}_{n,\sigma}, \quad \hat{D}_{n,\sigma}=\sum_{\boldsymbol{d} \in \Lambda} W_{n,\sigma}(\boldsymbol{d}) \hat{t}_{\sigma}(\boldsymbol{d}),\\
     &W_{n,\uparrow}(\boldsymbol{d})=W^*_{n,\downarrow}(\boldsymbol{d})=\eta_{\boldsymbol{d}} \bar{d}^{n} e^{-\frac{\pi|d|^{2}}{2}},
\end{aligned}
\end{equation}
where $\eta_{\boldsymbol{d}}=(-1)^{d_x+d_y+d_xd_y}$ and $d=\frac{1}{\sqrt{2}}(d_x+id_y)$ is the complex coordinate of the lattice site $\boldsymbol{d}$.  The hopping operator $\hat{t}_{\sigma}(\boldsymbol{d})$ takes: $\hat{t}_{\sigma}(\boldsymbol{d}) = e^{-i\sigma\frac{\pi}{2} d_x  d_y } \sum_{\boldsymbol{r} \in \Lambda} e^{i\pi d_x  r_y }|\boldsymbol{r}-\boldsymbol{d},\sigma\rangle\langle\boldsymbol{r},\sigma|$. Here we only consider an even LL-index $n$, since the spectrum of the odd-$n$ generalized KM model has symmetry protected band touching points. When $n=0$, the Bloch wave functions $|u(\mathbf{k})\rangle$ are also the exact ground states of the KM model discussed in the maintext.

Now we can directly verify that the phase factor  $\theta_{\mathbf{k}}$ is exactly zero for this class of wave functions:
\begin{equation}
\begin{aligned}
   \psi_{n,\mathbf{k}-\frac{\mathbf{Q}}{2}}(0) &=  \sum_{m \in \mathbb{Z}} H_{n}\left(k_{x} \ell_B-\frac{\pi l_B}{2a_x}+\left(2 \pi \ell_B / a_{x}\right) m\right) e^{-(1 / 2)\left(k_{x} \ell_B-\frac{\pi l_B}{2a_x}+\left(2 \pi \ell_B / a_{x}\right) m\right)^{2}} e^{-2i k_{y} a_{y} m}\\
    &=e^{-2ik_ya_y}\sum_{m \in \mathbb{Z}} H_{n}\left(k_{x} \ell_B+\frac{3\pi l_B}{2a_x}+\left(2 \pi \ell_B / a_{x}\right) m\right) e^{-(1 / 2)\left(k_{x} \ell_B+\frac{3\pi l_B}{2a_x}+\left(2 \pi \ell_B / a_{x}\right) m\right)^{2}} e^{-2i k_{y} a_{y} m}\\
    &=e^{-2ik_ya_y}\psi_{n,\mathbf{k}+\frac{\mathbf{Q}}{2}}(y=a_y) ,
\end{aligned}
\end{equation}
where in the second equality we shift the dummy index $m\rightarrow m+1$ and in the final equality we make use of the identity $\frac{a_y}{l_B}=\frac{a_yl_B}{l_B^2}=\frac{a_yl_B}{2a_xa_y/(2\pi)}=\frac{\pi l_B}{a_x}$.

Similarly, we have:
\begin{equation}
    \begin{aligned}
    \psi_{n,\mathbf{k}-\frac{\mathbf{Q}}{2}}(y=a_y) &=  \sum_{m \in \mathbb{Z}} H_{n}\left(k_{x} \ell_B+\frac{\pi l_B}{2a_x}+\left(2 \pi \ell_B / a_{x}\right) m\right) e^{-(1 / 2)\left(k_{x} \ell_B+\frac{\pi l_B}{2a_x}+\left(2 \pi \ell_B / a_{x}\right) m\right)^{2}} e^{-2i k_{y} a_{y} m}\\
    &=\psi_{n,\mathbf{k}+\frac{\mathbf{Q}}{2}}(0).
    \end{aligned}
\end{equation}
Hence we have proved that $\theta_{\mathbf{k}}=0$ in the generalized KM model. Generally, the phase $\theta_{\mathbf{k}}\neq0$ if we move away from the KM or generalized KM point by truncating the hopping but still preserve the MTS and TRS. In the following, we will keep the phase factor $\theta_{\mathbf{k}}$ throughout the derivation, so that the properties derived universally apply to gapped $\pi$-flux systems with time-reversal symmetry.

The order parameter we construct takes :
\begin{equation}
    \hat{\eta}^{\dagger}=\sum_{\mathbf{R}\in \text{unit cell}}(-1)^{R_x}\left[\hat{c}^{\dagger}(\mathbf{R}+\mathbf{a}_y)-\hat{c}^{\dagger}(\mathbf{R}-\mathbf{a}_y)\right]\sigma_x\hat{c}^{\dagger}(\mathbf{R}) =\sum_{\mathbf{k}}\hat{\gamma}_{m,\sigma}^{\dagger}(\mathbf{k}+\mathbf{Q}/2)F_{mn;\sigma\sigma^{\prime}}(\mathbf{k})(\hat{\gamma}_{n,\sigma^{\prime}}^{\dagger}(-\mathbf{k}+\mathbf{Q}/2))^{T},
\end{equation}
where $m,n$ are the energy band indices and $m=1,2$ labels the lowest flat band and the upper dispersive bands respectively. Since we focus on the $S_z=0$ pairing, all the same spin form factors $F_{mn;\sigma\sigma}(\mathbf{k})=0$ and we only need to consider the opposite-spin components. The off-diagonal (band-index) components of $F_{mn;\sigma\bar{\sigma}}(\mathbf{k})$ take:
\begin{equation}
    \begin{aligned}
        F_{12;\uparrow\downarrow}(\vec{k})&=\sum_{\mu\nu}(U^{\uparrow}_{\mu 1}(\vec{k}+\frac{\vec{Q}}{2}))^*N_{\mu\nu}(\vec{k})(U^{\downarrow}_{\nu 2}(\frac{\vec{Q}}{2}-\vec{k}))^*\\
        &=\sum_{\mu\nu}(U^{\uparrow}_{\mu 1}(\vec{k}+\frac{\vec{Q}}{2}))^*N_{\mu\nu}(\vec{k})U^{\uparrow}_{\nu2}(-\frac{\vec{Q}}{2}+\vec{k})\\
        &=\left(a^*_n(\vec{k}+\frac{\vec{Q}}{2}),b^*_n(\vec{k}+\frac{\vec{Q}}{2})\right)N_{\mu\nu}(\vec{k})\begin{pmatrix}
            -b^*_{n}(\vec{k}-\frac{\vec{Q}}{2})\\
            a^*_n(\vec{k}-\frac{\vec{Q}}{2})
        \end{pmatrix}\\
        &=(-2i\sin(k_y))e^{ik_y-i\theta_\mathbf{k}}\left(a^*_n(\vec{k}+\frac{\vec{Q}}{2}),b^*_n(\vec{k}+\frac{\vec{Q}}{2})\right)\tau_x\begin{pmatrix}
            -a^*_{n}(\vec{k}+\frac{\vec{Q}}{2})\\
            b^*_n(\vec{k}+\frac{\vec{Q}}{2})
        \end{pmatrix}\\
             &=0
    \end{aligned}
\end{equation}
where $N(\mathbf{k})=-2i\sin(k_y)\tau_x$ is the form factor of the order parameter in the sublattice space $\hat{\eta}^{\dagger}=\sum_{\mathbf{k}}\hat{c}^{\dagger}(\mathbf{k}+\frac{\mathbf{Q}}{2})N(\mathbf{k})(\hat{c}^{\dagger}(-\mathbf{k}+\frac{\mathbf{Q}}{2}))^T$ and the $\tau_x$ matrix acts on the sublattice index. Similarly, we can prove that $F_{12;\sigma\bar{\sigma}}=F_{21;\sigma\bar{\sigma}}=0$, where $\bar{\sigma}=-\sigma$ and hence the generalized QGN is satisfied.
\subsubsection{Off-diagonal long-range order of the generalized $\eta$-pairing state}
Here we follow the method in \cite{leggett2006quantum} to prove that the PDW state in the generalized $\eta$-pairing form has off-diagonal long-range order (ODLRO). First of all, the PDW state is: $|2N\rangle=\left(\hat{P}\eta^{\dagger}\hat{P}\right)^N|\text{vac}\rangle,\quad \hat{P}\eta^{\dagger}\hat{P}=-2i\sum_{\mathbf{k}}\left(\sin(k_y)e^{-ik_y+i\theta_\mathbf{k}}\right)\hat{\gamma}_{\uparrow}^{\dagger}(\mathbf{k}+\frac{\mathbf{Q}}{2})\hat{\gamma}_{\downarrow}^{\dagger}(-\mathbf{k}+\frac{\mathbf{Q}}{2}),$ where the model dependent phase factor $e^{i\theta_\mathbf{k}}=1$ at the KM point. The ODLRO of $|2N\rangle$ is equivalent to the pair-pair correlation: $\rho_2(\mathbf{R}_1,\mathbf{R}_2;\mathbf{R}_1^{\prime},\mathbf{R}_2^{\prime})=\frac{\langle 2N|\hat{\gamma}^{\dagger}_{\uparrow}(\mathbf{R}_1)\hat{\gamma}^{\dagger}_{\downarrow}(\mathbf{R}_2)\hat{\gamma}_{\downarrow}(\mathbf{R}_2^{\prime})\hat{\gamma}_{\uparrow}(\mathbf{R}_1^{\prime})|2N\rangle}{\langle 2N|2N\rangle}\neq 0$ in the limit $|\mathbf{R}_1-\mathbf{R}^{\prime}_2|\rightarrow +\infty$ while $|\mathbf{R}_1-\mathbf{R}_2|$ and $|\mathbf{R}^{\prime}_1-\mathbf{R}^{\prime}_2|$ are finite, and $\hat{\gamma}_{\sigma}^{\dagger}(\mathbf{R})=\frac{1}{\sqrt{S}}\sum_{\mathbf{k}}\hat{\gamma}_{\sigma}^{\dagger}(\mathbf{k})e^{-i\mathbf{k}\cdot\mathbf{R}}$ where $S$ is the number of unit cells. This correlation is the same as that of a BCS state $|\psi_{\text{BCS}}\rangle$ if we neglect corrections of relative order $N^{-\frac{1}{2}
}$ or smaller \cite{leggett2006quantum}, where the BCS state is:
\begin{equation}
    |\psi_{\text{BCS}}\rangle=N_{\text{BCS}}e^{\hat{\Omega}}|\text{vac}\rangle,\quad \hat{\Omega}=\lambda\sum_{\mathbf{k}}\left(\sin(k_y)e^{-ik_y+i\theta_\mathbf{k}}\right)\hat{\gamma}_{\uparrow}^{\dagger}(\mathbf{k}+\frac{\mathbf{Q}}{2})\hat{\gamma}_{\downarrow}^{\dagger}(-\mathbf{k}+\frac{\mathbf{Q}}{2}),\quad \sum_{\mathbf{k}}\frac{|\lambda|^2\sin^2(k_y)}{1+|\lambda|^2\sin^2(k_y)}=N,
\end{equation}
where $N_{\text{BCS}}$ is the normalization factor and  the parameter $\lambda$ is chosen such that the average electron number of the BCS state is $2N$.  The parameter $\lambda$ can always be found for $\nu=\frac{N}{S}\in (0,1)$, since $\sum_{\mathbf{k}}\frac{|\lambda|^2\sin^2(k_y)}{1+|\lambda|^2\sin^2(k_y)}= S (1-\frac{1}{\sqrt{1+\lambda^2}})$. Here we can see that $\lambda=0,\infty$ when $N/S=0$ or 1, and these two special fillings do not possess ODLRO, since they correspond to a vacuum state and fully occupied flat bands respectively.  Written explicitly in the momentum space, the pair-pair correlation is:
\begin{equation}
\begin{aligned}
    &\rho_2(\mathbf{R}_1,\mathbf{R}_2;\mathbf{R}_1^{\prime},\mathbf{R}_2^{\prime})=\langle \psi_{\text{BCS}}|\hat{\gamma}^{\dagger}_{\uparrow}(\mathbf{R}_1)\hat{\gamma}^{\dagger}_{\downarrow}(\mathbf{R}_2)\hat{\gamma}_{\downarrow}(\mathbf{R}_2^{\prime})\hat{\gamma}_{\uparrow}(\mathbf{R}_1^{\prime})|\psi_{\text{BCS}}\rangle\\
    &=\frac{1}{S^2 }[\sum_{\mathbf{k}_1,\mathbf{k}_2}e^{i\mathbf{k}_1\cdot(\mathbf{R}_2-\mathbf{R}_1)+i\mathbf{k}_2\cdot(\mathbf{R}^{\prime}_1-\mathbf{R}^{\prime}_2)}e^{-i\mathbf{Q}\cdot\frac{\mathbf{R}_1+\mathbf{R}_2}{2}}e^{i\mathbf{Q}\cdot\frac{\mathbf{R}^{\prime}_1+\mathbf{R}^{\prime}_2}{2}}\langle \psi_{\text{BCS}}|\hat{\gamma}^{\dagger}_{\uparrow}(\mathbf{k}_1+\frac{\mathbf{Q}}{2})\hat{\gamma}^{\dagger}_{\downarrow}(-\mathbf{k}_1+\frac{\mathbf{Q}}{2})\hat{\gamma}_{\downarrow}(-\mathbf{k}_2+\frac{\mathbf{Q}}{2})\hat{\gamma}_{\uparrow}(\mathbf{k}_2+\frac{\mathbf{Q}}{2})|\psi_{\text{BCS}}\rangle\\
    &+\sum_{\mathbf{k}_1,\mathbf{k}_2}e^{i\mathbf{k}_1\cdot(\mathbf{R}^{\prime}_1-\mathbf{R}_1)+i\mathbf{k}_2\cdot(\mathbf{R}^{\prime}_2-\mathbf{R}_2)}\langle \psi_{\text{BCS}}|\hat{\gamma}^{\dagger}_{\uparrow}(\mathbf{k}_1)\hat{\gamma}^{\dagger}_{\downarrow}(\mathbf{k}_2)\hat{\gamma}_{\downarrow}(\mathbf{k}_2)\hat{\gamma}_{\uparrow}(\mathbf{k}_1)|\psi_{\text{BCS}}\rangle].
\end{aligned}
\end{equation}
In the limit $|\mathbf{R}^{\prime}_1-\mathbf{R}_1|\rightarrow\infty,|\mathbf{R}^{\prime}_2-\mathbf{R}_2|\rightarrow\infty$, the second term is negligible due to the fast-oscillating phase factor. As a result, the pair-pair correlation is:
\begin{equation}
\begin{aligned}
       \rho_2(\mathbf{R}_1,\mathbf{R}_2;\mathbf{R}_1^{\prime},\mathbf{R}_2^{\prime})&=\frac{1}{S^2 }\sum_{\mathbf{k},\mathbf{k}^{\prime}}\left(e^{i\mathbf{k}\cdot(\mathbf{R}_2-\mathbf{R}_1)}e^{-i\mathbf{Q}\cdot\frac{\mathbf{R}_1+\mathbf{R}_2}{2}}\frac{\lambda \sin(k_y)e^{ik_y-i\theta_{\mathbf{k}}}}{1+\lambda^2\sin^2(k_y)}\right)\left(e^{i\mathbf{k}^{\prime}\cdot(\mathbf{R}^{\prime}_1-\mathbf{R}^{\prime}_2)}e^{i\mathbf{Q}\cdot\frac{\mathbf{R}^{\prime}_1+\mathbf{R}^{\prime}_2}{2}}\frac{\lambda \sin(k^{\prime}_y)e^{-ik^{\prime}_y+i\theta_{\mathbf{k}^{\prime}}}}{1+\lambda^2\sin^2(k^{\prime}_y)}\right)\\
       &=xSF(\mathbf{R}_1,\mathbf{R}_2)F^*(\mathbf{R}^{\prime}_1,\mathbf{R}^{\prime}_2),
\end{aligned}
\end{equation}
where $F(\mathbf{R}_1,\mathbf{R}_2)=\left(\frac{1}{\sqrt{S}}e^{-i\mathbf{Q}\cdot\frac{\mathbf{R}_1+\mathbf{R}_2}{2}}\right)f(\mathbf{R}_2-\mathbf{R}_1),\quad f(\mathbf{R}_2-\mathbf{R}_1)=\frac{1}{\sqrt{x}S}\sum_{\mathbf{k}}e^{i\mathbf{k}\cdot(\mathbf{R}_2-\mathbf{R}_1)}\frac{\lambda \sin(k_y)e^{ik_y-i\theta_{\mathbf{k}}}}{1+\lambda^2\sin^2(k_y)}$ and $x$ is the normalization factor such that $\sum_{\mathbf{R}}|f(\mathbf{R})|^2=\frac{1}{xS}\sum_{\mathbf{k}}(\frac{\lambda \sin(k_y)}{1+\lambda^2\sin^2(k_y)})^2=\frac{1}{x}\frac{\lambda^2}{2(1+\lambda^2)^{\frac{3}{2}}}=1$. Now it is obvious that the two-particle reduced density matrix $\rho_2(\mathbf{R}_1,\mathbf{R}_2;\mathbf{R}_1^{\prime},\mathbf{R}_2^{\prime})$ has a macroscopic eigenvalue $xS$ and $ \rho_2(\mathbf{R}_1,\mathbf{R}_2;\mathbf{R}_1^{\prime},\mathbf{R}_2^{\prime})\neq 0$ in the limit $|\frac{\mathbf{R}_1+\mathbf{R}_2}{2}-\frac{\mathbf{R}^{\prime}_1+\mathbf{R}^{\prime}_2}{2}|\rightarrow\infty$, which are two equivalent definitions of ODLRO.

\subsubsection{Triviality of the quadratic term }
The projected interaction $\hat{P}\hat{V}\hat{P}$ differs from $U\sum_{\mathbf{R},I}(\hat{\bar{S}}_{\mathbf{R}}^{(I)})^{\dagger}\hat{\bar{S}}_{\mathbf{R}}^{(I)}$ by a quadratic Hartree-Fock (HF) type term \cite{PhysRevX.14.041004}. This HF type term comes from the fact that the interaction is not in the normal order. Written explicitly, if  we denote the lattice fermion operator as: $\hat{c}=\bar{c}+\tilde{c}$, where $\bar{c}$ is the projected fermion operator which only involves the flat band operators $\hat{\gamma}_{1\sigma}(\mathbf{k})$, and $\tilde{c}$ is the remaining part, then we have:
\begin{equation}
    \hat{P}\hat{c}_{1}^{\dagger} \hat{c}_{2} \hat{c}_{3}^{\dagger} \hat{c}_{4}\hat{P} = \bar{c}_{1}^{\dagger} \bar{c}_{2} \bar{c}_{3}^{\dagger} \bar{c}_{4}+\langle \tilde{c}_{2} \tilde{c}_{3}^{\dagger}\rangle\bar{c}_{1}^{\dagger} \bar{c}_{4}+(\text { constant }),
\end{equation}
where $\langle \tilde{c}_{2} \tilde{c}_{3}^{\dagger}\rangle$ is the expectation value in empty upper bands. Here we only retain one term $\langle \tilde{c}_{2} \tilde{c}_{3}^{\dagger}\rangle\bar{c}_{1}^{\dagger} \bar{c}_{4}$, since all the other Wick contractions are zero, as the flat band is at the bottom of the spectrum.

In the model here, we prove that the Hartree-Fock type term is actually a chemical potential term in the flat band subspace and hence can be neglected. We begin with the Fourier transform of the local operators $\hat{S}^{(I)}_{\mathbf{q}}=\frac{1}{S}\sum_{\mathbf{R}}e^{i\mathbf{q}\cdot\mathbf{R}}\hat{S}^{(I)}_{\mathbf{R}}=\frac{1}{S}\sum_{\mathbf{k}}\hat{c}^{\dagger}(\mathbf{k}+\mathbf{q})S^{(I)}(\mathbf{k}+\mathbf{q},\mathbf{k})\hat{c}(\mathbf{k})$. The  form factors $S^{(I)}(\mathbf{k}+\mathbf{q},\mathbf{k})$ here are:
\begin{equation}
    S^{(1)}(\mathbf{k}+\mathbf{q},\mathbf{k})=-2i\sin(k_y+q_y)(\frac{\tau_x-i\tau_y}{2}),\quad   S^{(2)}(\mathbf{k}+\mathbf{q},\mathbf{k})=-2i\sin(k_y+q_y)e^{-iq_ya_y}(\frac{\tau_x+i\tau_y}{2}).
    \label{pi_form}
\end{equation}

The Hartree-Fock type term is actually:
\begin{equation}
\begin{aligned}
 \Delta H_{HF}&=\frac{1}{S}\sum_{\mathbf{k},\mathbf{q};m,n\in\text{flat};I=1,2}\sum_{m^{\prime}\notin \text{flat},\sigma}\gamma_{m,\sigma}^{\dagger}(\mathbf{k})(\tilde{S}^{(I)}(\mathbf{k},\mathbf{k}-\mathbf{q}))_{mm^{\prime};\sigma\sigma}^{\dagger}\tilde{S}_{m^{\prime}n;\sigma\sigma}^{(I)}(\mathbf{k},\mathbf{k}-\mathbf{q})\gamma_{n,\sigma}(\mathbf{k})\\
 &=\frac{1}{S}\sum_{\mathbf{k};m,n\in\text{flat}}\gamma_m^{\dagger}(\mathbf{k})\left\{U^{\dagger}(\mathbf{k})\left[\sum_{\mathbf{q},I}(S^{(I)}(\mathbf{k},\mathbf{q}))^{\dagger}Q(\mathbf{q})S^{(I)}(\mathbf{k},\mathbf{q}))\right]U(\mathbf{k})\right\}_{mn}\gamma_{n}(\mathbf{k}),
\end{aligned}
\end{equation}
where $\tilde{S}_{m^{\prime}n;\sigma\sigma^{\prime}}^{(I)}(\mathbf{k},\mathbf{k}-\mathbf{q})$ denote band-basis form factors corresponding to the sublattice-basis form factors $S^{(I)}(\mathbf{k},\mathbf{q})$ appearing in the square brackets, and $\langle\hat{\gamma}_{2,\sigma}(\mathbf{k})\hat{\gamma}_{2,\sigma^{\prime}}^{\dagger}(\mathbf{k})\rangle=\delta_{\sigma,\sigma^{\prime}}$, where $\hat{\gamma}_{2,\sigma}(\mathbf{k})$ are upper dispersive band operators; in the second line we have replaced the momentum $\mathbf{q}$ with $-\mathbf{q}+\mathbf{k}$.   $Q_{\sigma\sigma^{\prime}}(\mathbf{p})=\delta_{\sigma\sigma^{\prime}}Q_{\sigma}(\mathbf{p})$ is the projection to the upper band at momentum $\mathbf{q}$: 
\begin{equation}
\begin{aligned}
     Q_{\uparrow}(\mathbf{p})&=\begin{pmatrix}
        |b_{n}(\mathbf{q})|^2&-a_n(\mathbf{q})b^*_n(\mathbf{\mathbf{q}})\\
        -a^*_n(\mathbf{q})b_n(\mathbf{\mathbf{q}})&|a_{n}(\mathbf{q})|^2
    \end{pmatrix},\\
    Q_{\downarrow}(\mathbf{p})&=(Q_{\uparrow}(-\mathbf{p}))^*.
\end{aligned}
\end{equation}
Hence the Hartree-Fock type term can be simplified as:
\begin{equation}
    \begin{aligned}
        \Delta H_{HF}&=\frac{1}{2S}\sum_{\mathbf{k};m,n\in\text{flat}}\gamma_m^{\dagger}(\mathbf{k})\left\{U^{\dagger}(\mathbf{k})\left[\sum_{\mathbf{q}}(2\sin(q_y))^2\left(\tau_xQ(\mathbf{q})\tau_x+\tau_yQ(\mathbf{q})\tau_y\right)\right]U(\mathbf{k})\right\}_{mn}\gamma_{n}(\mathbf{k})
    \end{aligned}
    \label{hfterm}
\end{equation}

Now it is clear that $\tau_xQ(\mathbf{q})\tau_x+\tau_yQ(\mathbf{q})\tau_y$ is a diagonal matrix:
\begin{equation}
   \tau_xQ(\mathbf{q})\tau_x+\tau_yQ(\mathbf{q})\tau_y=\begin{pmatrix}
  |a_n(\mathbf{q})|^2&0 \\
        0&|b_n(\mathbf{q})|^2
    \end{pmatrix}\otimes\sigma_0,
\end{equation}
where we have replaced $\mathbf{p}\rightarrow-\mathbf{p}$ in the spin-down subspace. As a result, we can reach the conclusion that the factor in the 
brackets in Eq.~\eqref{hfterm} is proportional to the identity matrix after the integration of the momentum $\mathbf{q}$, since we have $a_n(\mathbf{q})=e^{-iq_y+i\theta_{\mathbf{q}}}b_n(\mathbf{q}+\mathbf{Q}),b_n(\mathbf{q})=e^{-iq_y+i\theta_{\mathbf{q}}}a_n(\mathbf{q}+\mathbf{Q})$ with $\mathbf{Q}=(\pi,0)$.  As a result, the quadratic HF type term $ \Delta H_{HF}$ is a chemical potential term in the flat band space and can be neglected.
\subsection{ $A_2$. $\phi=\frac{1}{p}$-flux systems}
\subsubsection{Generalized QGN}
Here we still use the MTS to prove that the PDW order parameter $\hat{\eta}_p^{\dagger}$ satisfies the generalized QGN.  We take the magnetic unit cell as: $\tilde{a}_x=a_x,\tilde{a}_y=pa_y$, and the flat band Bloch wave functions are:
\begin{equation}
    |\Phi^{(m)}_{\mathbf{k},\uparrow}\rangle=\sum_{\mu=0}^{p-1}a^{(m)}_{\mu}(\mathbf{k})|\mathbf{k},\mu,\uparrow\rangle, |\Phi^{(m)}_{\mathbf{k},\downarrow}\rangle=\sum_{\mu=0}^{p-1}(a^{(m)}_{\mu}(-\mathbf{k}))^*|\mathbf{k},\mu,\downarrow\rangle,
\end{equation}
 where $m$ is the energy band label, and we assume $m=1$ labels the flat band. $|\mathbf{k},\mu,\sigma\rangle=\frac{1}{\sqrt{S}}\sum_{\mathbf{R}}e^{i\mathbf{k}\cdot(\mathbf{R}+\mu a_y)}|\mathbf{R}+\mu\mathbf{a}_y,\sigma\rangle$ and $\mu\mathbf{a}_y$ is the intra-unit cell coordinate. Here we assume $|\Phi^{(m)}_{\mathbf{k},\sigma}\rangle$ is normalized. Now we can arrive at the following properties of Bloch functions using magnetic translation symmetries:

 \begin{equation}
 \begin{aligned}
       (\hat{\tilde{T}}_{a_y})^{-1}|\Phi^{(m)}_{\mathbf{k},\sigma}\rangle\propto |\Phi^{(m)}_{\mathbf{k}+\sigma\mathbf{Q}_p,\sigma}\rangle,& \Longrightarrow a^{(m)}_{\mu}(\mathbf{k})=e^{ik_ya_y+i\theta^{(m)}_{\mathbf{k}}} a^{(m)}_{\mu+1}(\mathbf{k}-\mathbf{Q}_p),
 \end{aligned}
 \end{equation}
 where $\mathbf{Q}_p=(\frac{2\pi}{p},0)$, the index $\mu$ of $a^{(m)}_\mu(\mathbf{k})$ are always mod $p$, and $e^{i\theta^{(m)}_{\mathbf{k}}}$ is a model-dependent phase factor. It  is now direct to verify the off-diagonal elements of  $F_{nm;\sigma\sigma^{\prime}}(\mathbf{k})$ in the order parameter $\eta_p^{\dagger}=\sum_{\mathbf{r}}e^{i\mathbf{Q}_p\cdot\mathbf{r}}\hat{c}_{\downarrow}^{\dagger}(\mathbf{r}+\mathbf{a}_y)\hat{c}_{\uparrow}^{\dagger}(\mathbf{r})=\sum_{\mathbf{k}}F_{nm;\sigma\sigma^{\prime}}(\mathbf{k})\hat{\gamma}^{\dagger}_{n,\sigma}(\mathbf{k}+\frac{\mathbf{Q}_p}{2})\hat{\gamma}^{\dagger}_{m,\sigma^{\prime}}(-\mathbf{k}+\frac{\mathbf{Q}_p}{2})$ in the energy band basis are zero: 
 \begin{equation}
 \begin{aligned}
      F_{m1;\downarrow\uparrow}(\mathbf{k})&=\frac{1}{2}e^{-ik_y}\sum_{\mu}(a_{\mu}^{(m)}(-\mathbf{k}+\frac{\mathbf{Q}_p}{2}))^*a_{\mu+1}^{(1)}(-\mathbf{k}-\frac{\mathbf{Q}_p}{2})=\frac{1}{2}e^{-i\theta^{(1)}_{-\mathbf{k}+\frac{\mathbf{Q}_p}{2}}}\sum_{\mu}(a_{\mu}^{(m)}(-\mathbf{k}+\frac{\mathbf{Q}_p}{2}))^*a_{\mu}^{(1)}(-\mathbf{k}+\frac{\mathbf{Q}_p}{2})=0,\\
      F_{m1;\uparrow\downarrow}(\mathbf{k})&=-\frac{1}{2}e^{ik_y}\sum_{\mu}(a_{\mu}^{(m)}(\mathbf{k}+\frac{\mathbf{Q}_p}{2}))^*a_{\mu+1}^{(1)}(\mathbf{k}-\frac{\mathbf{Q}_p}{2})=-\frac{1}{2}e^{-i\theta^{(1)}_{\mathbf{k}+\frac{\mathbf{Q}_p}{2}}}\sum_{\mu}(a_{\mu}^{(m)}(\mathbf{k}+\frac{\mathbf{Q}_p}{2}))^*a_{\mu}^{(1)}(\mathbf{k}+\frac{\mathbf{Q}_p}{2})=0,
 \end{aligned}
 \end{equation}
where $m>1$ label the non-flat bands and similarly we can prove $F_{1m;\uparrow\downarrow}(\mathbf{k})= F_{1m;\downarrow\uparrow}(\mathbf{k})=0$. As a result, the order parameter $\hat{\eta}_p^{\dagger}$ satisfies the generalized QGN.

  The projected order parameter takes a simple form:
 \begin{equation}
     \begin{aligned}
         \hat{P}\hat{\eta}_p^{\dagger}\hat{P}&=\sum_{\mathbf{k}}\hat{\gamma}^{\dagger}_{1,\downarrow}(\mathbf{k}+\frac{\mathbf{Q}_p}{2})\hat{\gamma}^{\dagger}_{1,\uparrow}(-\mathbf{k}+\frac{\mathbf{Q}_p}{2})e^{-i\theta^{(1)}_{-\mathbf{k}+\frac{\mathbf{Q}_p}{2}}}.
     \end{aligned}
 \end{equation}
\subsubsection{Triviality of the quadratic term}
The Hartree-Fock type term arising from the projection can be proved to be a chemical potential term similar to the $\pi$-flux model. We begin with the Fourier transformation $\hat{S}^{(\mu)}_{\mathbf{q}}=\frac{1}{S}\sum_{\mathbf{R}}e^{i\mathbf{q}\cdot\mathbf{R}}\hat{S}^{(\mu)}_{\mathbf{R}}=\frac{1}{S}\sum_{\mathbf{k},\sigma}\hat{c}_{\sigma}^{\dagger}(\mathbf{k}+\mathbf{q})S^{(\mu)}_{\sigma}(\mathbf{k}+\mathbf{q},\mathbf{k})\hat{c}_{\sigma}(\mathbf{k})$. The  form factors $S_{\sigma}^{(m)}(\mathbf{k}+\mathbf{q},\mathbf{k})$ here are:
\begin{equation}
    S_{\uparrow}^{(\mu)}(\mathbf{k}+\mathbf{q},\mathbf{k})=e^{-i k_ya_y-i\mu q_ya_y}(\frac{\tau^{(\mu,\mu-1)}_x+i\tau^{(\mu,\mu-1)}_y}{2}),S_{\downarrow}^{(\mu)}(\mathbf{k}+\mathbf{q},\mathbf{k})=-e^{i k_ya_y-i\mu q_ya_y}(\frac{\tau^{(\mu,\mu+1)}_x+i\tau^{(\mu,\mu+1)}_y}{2}),
\end{equation}
where $\tau_{x,y}^{(\mu,\mu-1)}$ are the Pauli matrices in the sublattice space $(\hat{c}_{\mu,\sigma}(\mathbf{k}),\hat{c}_{\mu-1,\sigma}(\mathbf{k}))^T$, and all the other elements are zero. Here the sublattice indices of the Pauli matrices are always mod p.
The Hartree-Fock type term here takes:
\begin{equation}
\begin{aligned}
 \Delta H_{HF} &=\frac{1}{S}\sum_{\mathbf{k};m,n\in\text{flat}}\gamma_m^{\dagger}(\mathbf{k})\left\{U^{\dagger}(\mathbf{k})\left[\sum_{\mathbf{q},\mu}S^{(\mu)}(\mathbf{k},\mathbf{q})Q(\mathbf{q})(S^{(\mu)}(\mathbf{k},\mathbf{q}))^{\dagger}+(S^{(\mu)}(\mathbf{k},\mathbf{q}))^{\dagger}Q(\mathbf{q})S^{(\mu)}(\mathbf{k},\mathbf{q})\right]U(\mathbf{k})\right\}_{mn}\gamma_{n}(\mathbf{k}),
\end{aligned}
\label{hartree_fock}
\end{equation}
where the labels $m,n$ include the spin and flat-band indices (here the flat band index can only take 1). The column $U_{\mu m}(\mathbf{k}),\mu=1,2...p$ of the unitary matrix $U(\mathbf{k})$ is the Bloch wavefunction of the band m. $Q(\mathbf{q})=\mathbb{I}-P(\mathbf{q})$ is the projection to non-flat bands in the sublattice basis, and $P(\mathbf{q})$ is the projection to the flat band.  As a result, the term in the bracket is in the form:
\begin{equation}
    \begin{aligned}
        \sum_{\mathbf{q},\mu}\left[\tau^{(\mu,\mu-1)}_xQ_{\uparrow}(\mathbf{q})\tau^{(\mu,\mu-1)}_x+\tau^{(\mu,\mu-1)}_yQ_{\uparrow}(\mathbf{q})\tau^{(\mu,\mu-1)}_y+\tau^{(\mu,\mu+1)}_xQ_{\downarrow}(\mathbf{q})\tau^{(\mu,\mu+1)}_x+\tau^{(\mu,\mu+1)}_yQ_{\downarrow}(\mathbf{q})\tau^{(\mu,\mu+1)}_y\right],
    \end{aligned}
\end{equation}
which is proportional to identity after the integration of $\mathbf{q}$ and the sum over $\mu$, since the Bloch wave functions satisfy $|a_{\mu}^{(1)}(\mathbf{q})|^2=|a_{\mu+1}^{(1)}(\mathbf{q}-\mathbf{Q}_p)|^2$ for any $\mu$.

\begin{table}[htbp]
\centering
\renewcommand{\arraystretch}{1.4}
\setlength{\tabcolsep}{7pt}
\begin{minipage}{0.48\textwidth}
\raggedright
\textbf{(a)}
\vspace{0.3em}
\makebox[\linewidth][c]{
\begin{tabular}{c|ccccccccc}
\diagbox[width=3em,height=2.2em,innerleftsep=1pt,innerrightsep=1pt]{$N_{\uparrow}$}{$N_{\downarrow}$}& 0 & 1 & 2 & 3 & 4 & 5 & 6 & 7 & 8 \\
\hline
0 & 1 & 0 & 0 & 0 & 1 & 4 & 6 & 4 & 1 \\
1 & 0 & 1 & 0 & 0 & 0 & 0 & 0 & 0 & 0 \\
2 & 0 & 0 & 1 & 0 & 0 & 0 & 0 & 0 & 0 \\
3 & 0 & 0 & 0 & 1 & 0 & 0 & 0 & 0 & 0 \\
4 & 1 & 0 & 0 & 0 & 1 & 4 & 6 & 4 & 1 \\
5 & 4 & 0 & 0 & 0 & 4 & 16 & 24 & 16 & 4 \\
6 & 6 & 0 & 0 & 0 & 6 & 24 & 36 & 24 & 6 \\
7 & 4 & 0 & 0 & 0 & 4 & 16 & 24 & 16 & 4 \\
8 & 1 & 0 & 0 & 0 & 1 & 4 & 6 & 4 & 1 \\
\end{tabular}
}
\end{minipage}
\begin{minipage}{0.48\textwidth}
\raggedright
\textbf{(b)}
\vspace{0.3em}
\makebox[\linewidth][c]{
\begin{tabular}{c|ccccccccc}
\diagbox[width=3em,height=2.2em,innerleftsep=1pt,innerrightsep=1pt]{$N_{\uparrow}$}{$N_{\downarrow}$} & 0 & 1 & 2 & 3 & 4 & 5 & 6 & 7 & 8 \\
\hline
0 & 1 & 0 & 0 & 0 & 1 & 0 & 0 & 0 & 0 \\
1 & 0 & 1 & 0 & 0 & 0 & 0 & 0 & 0 & 0 \\
2 & 0 & 0 & 1 & 0 & 0 & 0 & 0 & 0 & 0 \\
3 & 0 & 0 & 0 & 1 & 0 & 0 & 0 & 0 & 0 \\
4 & 1 & 0 & 0 & 0 & 1 & 0 & 0 & 0 & 0 \\
5 & 0 & 0 & 0 & 0 & 0 & 0 & 0 & 0 & 0 \\
6 & 0 & 0 & 0 & 0 & 0 & 0 & 0 & 0 & 0 \\
7 & 0 & 0 & 0 & 0 & 0 & 0 & 0 & 0 & 0 \\
8 & 0 & 0 & 0 & 0 & 0 & 0 & 0 & 0 & 0 \\
\end{tabular}
}
\end{minipage}

\caption{Zero-mode counting for the $\pi-$flux Kapit--Mueller model with the Landau level index $n=0$ on a $4\times4$ lattice from exact diagonalization,
equivalently $4\times2$ magnetic unit cells, with 8 allowed momenta in each
flat band. The row and column labels denote the number of spin-up and down electrons: $N_{\uparrow}$ and
$N_{\downarrow}$, respectively. Each entry gives the number of zero modes,
or equivalently the GSD, in the corresponding $(N_{\uparrow},N_{\downarrow})$
sector. The resulting GSDs at high filling are consistent with the number $\binom{L_x}{N_{\rm{tot}}-N_\uparrow}
\binom{L_x}{N_{\rm{tot}}-N_\downarrow}
$   in the region $N_{\uparrow,\downarrow}\geq N_{\rm{tot}}-L_x$, and $\binom{L_x}{N_{\rm{tot}}-N_{\uparrow/\downarrow}}$ in the two
 regions $N_{\uparrow/\downarrow}\geq N_{\rm{tot}}-L_x,N_{\downarrow/\uparrow}=0$, as discussed in the maintext. Panel (a) shows the original model, while panel (b) shows the result
after adding an energy-penalty term on the $k_y=0$ axis.}

\label{GSD}
\end{table}

\section{B. Exact Diagonalization and Families of Zero Modes}\label{sec:ed}

In this section, we present exact diagonalization (ED) results for several finite-size $\pi-$flux Kapit--Mueller model with the Landau level index $n=0$, and identify two families of zero modes. In Table~\ref{GSD}(a), we report the number of zero modes, equivalently the ground-state degeneracy (GSD), obtained from ED for a $4\times4$ system, corresponding to $4\times2$ magnetic unit cells, under periodic boundary conditions.
At first glance, there are multiple degenerate ground states at high fillings; nevertheless, as we show below, the fillings with multiple ground states  form a measure-zero set in the thermodynamic limit. 

\begin{figure}
    \centering
    \includegraphics[width=1.0\linewidth]{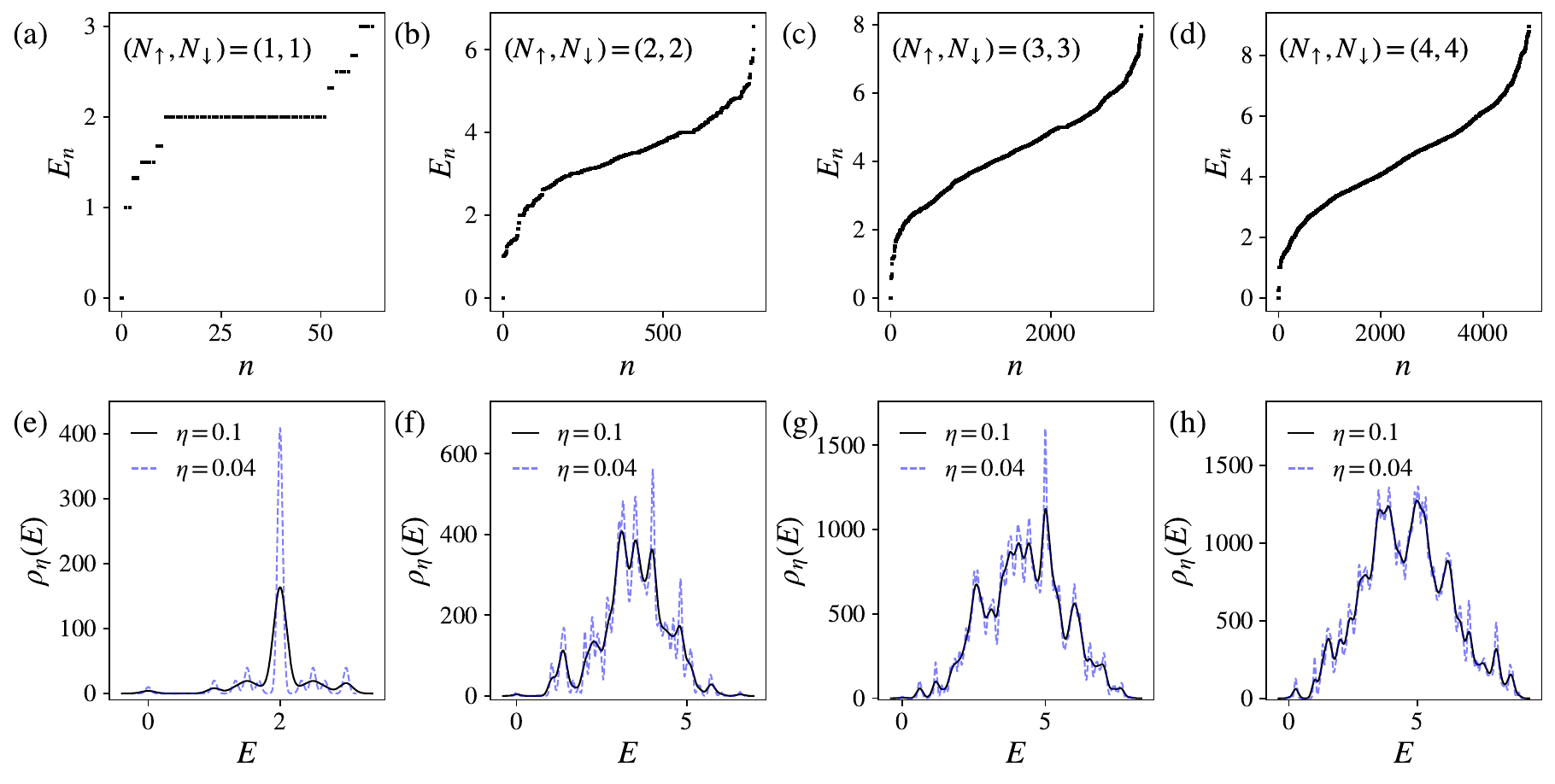}
    \caption{Energy spectra and density of states (DOS) in the $(N_\uparrow,N_\downarrow)=(1,1)$, $(2,2)$, $(3,3)$, and $(4,4)$ particle-number sectors for the $4\times 4$ system with $U=1$. 
    Panels (a)–(d) show the many-body eigenenergies $E_n$, ordered by increasing energy. 
    Panels (e)–(h) display the corresponding DOS obtained by Gaussian broadening of the discrete spectrum, $\rho_\eta(E)=\sum_n \frac{1}{\sqrt{2\pi}\eta}\exp\left(-\frac{(E-E_n)^2}{2\eta^2}\right).$
    Results are shown for $\eta=0.04$ (blue dashed lines) and $\eta=0.1$ (black solid lines). 
}
    \label{fig:S1}
\end{figure}

At low fillings, there is a unique zero mode when $N_\uparrow = N_\downarrow$, and no zero modes when $N_\uparrow \neq N_\downarrow$. These zero modes are precisely the $\eta$-pairing PDW states constructed in the maintext.
We now turn to the origin of the additional zero modes that appear at higher filling. The key observation is that, after projection, the local positive-semidefinite terms in the effective Hamiltonian acquire momentum-space form factors that vanish along a special line in the Brillouin zone. To make this structure explicit, we rewrite the projected Hamiltonian as (we neglect the trivial chemical potential term coming from the projection):
\begin{equation}
\begin{aligned}
        \hat{H}_{\text{eff}}&=U\sum_{\mathbf{R};I=1,2}(\hat{P}\hat{S}^{(I)}_{\mathbf{R}}\hat{P})^{\dagger}\left(\hat{P}\hat{S}^{(I)}_{\mathbf{R}}\hat{P}\right), U>0,\\
        \hat{P}\hat{S}^{(1),(2)}_{\mathbf{R}}\hat{P}&=\sum_{\mathbf{p},\mathbf{q}}e^{i\mathbf{R}\cdot(\mathbf{p}-\mathbf{q})}\hat{\gamma}^{\dagger}(\mathbf{p})S^{(1),(2)}(\mathbf{p},\mathbf{q})\hat{\gamma}(\mathbf{q}),\\
    S^{(1)}(\mathbf{p},\mathbf{q})&=\begin{pmatrix}
            b_0^*(\mathbf{p})a_0(\mathbf{q})(-2i\sin(p_y))&0\\
            0&b_0(-\mathbf{p})a_0^*(-\mathbf{q})(-2i\sin(p_y))
        \end{pmatrix},\\
        S^{(2)}(\mathbf{p},\mathbf{q})&=e^{-i(p_y-q_y)}\begin{pmatrix}
            a_0^*(\mathbf{p})b_0(\mathbf{q})(-2i\sin(p_y))&0\\
            0&a_0(-\mathbf{p})b_0^*(-\mathbf{q})(-2i\sin(p_y))
        \end{pmatrix}.
\end{aligned}
\label{pro_ham}
\end{equation}
Each term in the Fourier representation of $\hat{P}\hat{S}^{(I)}_{\mathbf{R}}\hat{P}$ contains a factor of $\sin(p_y)$, which singles out the $p_y=0$ axis. Consequently, if all momenta away from the $p_y=0$ axis are fully occupied, an arbitrary number of additional electrons may be placed on the $p_y=0$ axis. All states constructed in this way are annihilated by $\hat{P}\hat{S}^{(I)}_{\mathbf{R}}\hat{P}$ and are therefore ground states (zero modes). 

Further physical insight can be obtained from the projected PDW order parameter in Eq.~\eqref{eq:eta_pairing_form_factor}. Since $\hat{P}\eta^{\dagger}\hat{P}$ also contains a factor of $\sin(k_y)$, electrons on the $k_y=0$ axis do not participate in the pairing. Consequently, as the power $N$ is increased in
$|2N\rangle=(\hat{P}\eta^{\dagger}\hat{P})^N|\text{vac}\rangle$,
the states $|2N\rangle$ are initially the only ground states. When $N=L_xL_y/2-L_x$, all states away from the $p_y=0$ axis are fully occupied, so applying another $\eta^\dagger$ annihilates the state. Precisely at this point, additional ground states appear: the $k_y=0$ axis can be partially filled without changing the energy.
We can justify our above understanding by introducing an energy-penalty term on the $k_y=0$ modes: $\mu\sum_{k_y=0,\sigma}\hat{\gamma}_{\sigma}^{\dagger}(\mathbf{k})\hat{\gamma}_{\sigma}(\mathbf{k})$, $\mu>0$ and  repeating the ED calculation. The resulting GSD is illustrated in Table~\ref{GSD}(b). 
We find the degeneracy is lifted, leaving only the PDW state $|2N\rangle$ as the unique ground state. The only exceptions occur in the $(4,0), (0,4)$ sectors. These sectors correspond to the spin-polarized cases where the $k_y=0$ axis of the spin-up or spin-down band is completely empty, while all remaining states are fully occupied. Consequently, these states are not affected by the energy-penalty term.

We are therefore led to the following picture for the families of zero modes in the system. Let $M=\frac{L_xL_y}{2}-L_x$
denote the number of momenta away from the $k_y=0$ axis. For $N_\uparrow,N_\downarrow<M$, there is a zero mode only when $N_\uparrow=N_\downarrow=N$, in which case it is the unique $\eta$-pairing state $|2N\rangle=(\hat{P}\eta^{\dagger}\hat{P})^N|\text{vac}\rangle$.
When both $N_\uparrow$ and $N_\downarrow$ are greater than or equal to $M$, the zero modes are obtained by fully occupying all momenta away from the $k_y=0$ axis and choosing holes among the $L_x$ modes on the $k_y=0$ axis. 
More explicitly, let $|\mathrm{full},\sigma\rangle$ denote the state in which all momenta in the flat band are occupied by spin-$\sigma$ electrons. Then a zero mode in the $(N_\uparrow,N_\downarrow)$ sector can be constructed as
$\prod_{i=1}^{N_{\rm{tot}}-N_\uparrow}\hat\gamma_{\uparrow}(k_x^i,k_y=0)\prod_{j=1}^{N_{\rm{tot}}-N_\downarrow}\hat \gamma_{\downarrow}(k_x^j,k_y=0)\ket{\text{full},\uparrow}\otimes\ket{\text{full},\downarrow}$
where $N_{\rm tot}=L_xL_y/2$ is the total number of momenta in the magnetic Brillouin zone. Therefore, in the $(N_\uparrow,N_\downarrow)$ sector, the corresponding GSD is 
$
\binom{L_x}{N_{\rm{tot}}-N_\uparrow}
\binom{L_x}{N_{\rm{tot}}-N_\downarrow}
$.
Similar zero modes also appear in the fully spin-polarized sectors. For $N_\uparrow=0$ and $N_\downarrow\geq M$, or for $N_\downarrow=0$ and $N_\uparrow\geq M$, the GSD is respectively
$
\binom{L_x}{N_{\rm{tot}}-N_\downarrow}$ and $
\binom{L_x}{N_{\rm{tot}}-N_\uparrow}.
$

We further calculate the GSD for larger systems of sizes $6\times4$ and $4\times6$. The results, shown in Table~\ref{GSD_larger_system}, are consistent with the above discussion.
It remains to ask whether the special role of the $k_y=0$ axis has any significance in the thermodynamic limit. In general, the additional ground states occur only within the filling intervals $\nu\in [2-2/L,2]\cup [1-1/L,1]$,
which shrink to a measure-zero set as $L\rightarrow\infty$. We therefore conclude that, at generic electron filling, the PDW ground state is unique.

Moreover, in Fig.~\ref{fig:S1}, we also show the full many-body energy spectrum and the corresponding density of states for the $4\times4$ system in several particle-number sectors. In the filling regime with a PDW ground state, the low energy spectrum is relatively sparse and the peak of the density of states is roughly located at the middle of the whole spectrum.

\begin{table}[htbp]
\centering
\renewcommand{\arraystretch}{1.4}
\setlength{\tabcolsep}{4pt}
\begin{minipage}{0.48\textwidth}
\raggedright
\textbf{(a)}
\vspace{0.3em}
\makebox[\linewidth][c]{
\begin{tabular}{c|ccccccccccccc}
\diagbox[width=3em,height=2.2em,innerleftsep=1pt,innerrightsep=1pt]
{$N_{\uparrow}$}{$N_{\downarrow}$}
& 0 & 1 & 2 & 3 & 4 & 5 & 6 & 7 & 8 & 9 & 10 & 11 & 12 \\
\hline
0  & 1 & 0 & 0 & 0 & 0 & 0 & 0 & 0 & 1 & 4  & 6  & 4  & 1 \\
1  & 0 & 1 & 0 & 0 & 0 & 0 & 0 & 0 & 0 & 0  & 0  & 0  & 0 \\
2  & 0 & 0 & 1 & 0 & 0 & 0 & 0 & 0 & 0 & 0  & 0  & 0  & 0 \\
3  & 0 & 0 & 0 & 1 & 0 & 0 & 0 & 0 & 0 & 0  & 0  & 0  & 0 \\
4  & 0 & 0 & 0 & 0 & 1 & 0 & 0 & 0 & 0 & 0  & 0  & 0  & 0 \\
5  & 0 & 0 & 0 & 0 & 0 & 1 & 0 & 0 & 0 & 0  & 0  & 0  & 0 \\
6  & 0 & 0 & 0 & 0 & 0 & 0 & 1 & 0 & 0 & 0  & 0  & 0  & 0 \\
7  & 0 & 0 & 0 & 0 & 0 & 0 & 0 & 1 & 0 & 0  & 0  & 0  & 0 \\
8  & 1 & 0 & 0 & 0 & 0 & 0 & 0 & 0 & 1 & 4  & 6  & 4  & 1 \\
9  & 4 & 0 & 0 & 0 & 0 & 0 & 0 & 0 & 4 & 16 & 24 & 16 & 4 \\
10 & 6 & 0 & 0 & 0 & 0 & 0 & 0 & 0 & 6 & 24 & 36 & 24 & 6 \\
11 & 4 & 0 & 0 & 0 & 0 & 0 & 0 & 0 & 4 & 16 & 24 & 16 & 4 \\
12 & 1 & 0 & 0 & 0 & 0 & 0 & 0 & 0 & 1 & 4  & 6  & 4  & 1 \\
\end{tabular}
}
\end{minipage}
\begin{minipage}{0.48\textwidth}
\raggedright
\textbf{(b)}
\vspace{0.3em}
\makebox[\linewidth][c]{%
\begin{tabular}{c|ccccccccccccc}
\diagbox[width=3em,height=2.2em,innerleftsep=1pt,innerrightsep=1pt]
{$N_{\uparrow}$}{$N_{\downarrow}$}
& 0 & 1 & 2 & 3 & 4 & 5 & 6 & 7 & 8 & 9 & 10 & 11 & 12 \\
\hline
0  & 1  & 0 & 0 & 0 & 0 & 0 & 1  & 6   & 15  & 20  & 15  & 6  & 1 \\
1  & 0  & 1 & 0 & 0 & 0 & 0 & 0  & 0   & 0   & 0   & 0   & 0  & 0 \\
2  & 0  & 0 & 1 & 0 & 0 & 0 & 0  & 0   & 0   & 0   & 0   & 0  & 0 \\
3  & 0  & 0 & 0 & 1 & 0 & 0 & 0  & 0   & 0   & 0   & 0   & 0  & 0 \\
4  & 0  & 0 & 0 & 0 & 1 & 0 & 0  & 0   & 0   & 0   & 0   & 0  & 0 \\
5  & 0  & 0 & 0 & 0 & 0 & 1 & 0  & 0   & 0   & 0   & 0   & 0  & 0 \\
6  & 1  & 0 & 0 & 0 & 0 & 0 & 1  & 6   & 15  & 20  & 15  & 6  & 1 \\
7  & 6  & 0 & 0 & 0 & 0 & 0 & 6  & 36  & 90  & 120 & 90  & 36 & 6 \\
8  & 15 & 0 & 0 & 0 & 0 & 0 & 15 & 90  & 225 & 300 & 225 & 90 & 15 \\
9  & 20 & 0 & 0 & 0 & 0 & 0 & 20 & 120 & 300 & 400 & 300 & 120 & 20 \\
10 & 15 & 0 & 0 & 0 & 0 & 0 & 15 & 90  & 225 & 300 & 225 & 90 & 15 \\
11 & 6  & 0 & 0 & 0 & 0 & 0 & 6  & 36  & 90  & 120 & 90  & 36 & 6 \\
12 & 1  & 0 & 0 & 0 & 0 & 0 & 1  & 6   & 15  & 20  & 15  & 6  & 1 \\
\end{tabular}%
}
\end{minipage}

\caption{Zero-mode counting of the $\pi$-flux Kapit--Mueller model  with the Landau level index $n=0$ from ED calculations.
The row and column labels denote the numbers of spin-up and spin-down electrons,
$N_{\uparrow}$ and $N_{\downarrow}$, respectively. Each entry gives the number
of zero modes, or equivalently the GSD, in the corresponding
$(N_{\uparrow},N_{\downarrow})$ sector. Panels (a) and (b) correspond to the
$4\times6$ and $6\times4$ systems, respectively. 
For these two systems, we use sparse diagonalization with different Krylov-subspace dimensions to check convergence. The resulting GSDs at high filling are consistent with the number $\binom{L_x}{N_{\rm{tot}}-N_\uparrow}
\binom{L_x}{N_{\rm{tot}}-N_\downarrow}
$   in the region $N_{\uparrow,\downarrow}\geq N_{\rm{tot}}-L_x$, and $\binom{L_x}{N_{\rm{tot}}-N_{\uparrow/\downarrow}}$ in the two
 regions $N_{\uparrow/\downarrow}\geq N_{\rm{tot}}-L_x,N_{\downarrow/\uparrow}=0$, as discussed in the maintext.
}

\label{GSD_larger_system}
\end{table}

\section{C. Exactly solvable PDW of spinless fermions beyond QGN construction}
Here we construct exactly solvable PDW in spinless fermions using the ansatz  proposed in \cite{han2025}. We only require magnetic translation and inversion symmetries: the  system has $\pi$-flux per plaquette and is symmetric under inversion: $|\mathbf{r}\rangle\rightarrow|-\mathbf{r}\rangle$. A typical example is the spinless $\pi$-flux generalized Kapit-Mueller model \cite{shen2025} $\hat{H}_{\text{GKM},n}$ with an even Landau-level index $n$ as discussed before. Since we consider spinless fermions, there is only one flat band and we denote the projection to the flat band as $\hat{P}$. We still take the Landau gauge and the magnetic translation symmetries $\tilde{\hat{T}}_{a_x},\tilde{\hat{T}}_{2a_y}$ become the ordinary lattice translation symmetries. We still take the magnetic unit cell as $|\mathbf{R}=m\mathbf{a}_x+2k\mathbf{a}_y\rangle,|\mathbf{R}+\mathbf{a}_y\rangle,\quad m,k\in\mathbb{Z}$ which are labeled as A and B sublattices respectively. In these kind of models, the Bloch wave functions  can always be written as: $\rho_n(\mathbf{k})=a_n(\mathbf{k})|\mathbf{k},A\rangle+b_n(\mathbf{k})|\mathbf{k},B\rangle$, where $|a_n(\mathbf{k})|^2+|\mathbf{b}_n(\mathbf{k})|^2=1$. Due to the magnetic translation symmetry and inversion symmetry, the wave functions further satisfy:
\begin{equation}
\begin{aligned}
     & a_n(\mathbf{k}-\frac{\mathbf{Q}}{2})=e^{-ik_ya_y} e^{i\theta_n(\mathbf{k})}b_n(\mathbf{k}+\frac{\mathbf{Q}}{2}),b_n(\mathbf{k}-\frac{\mathbf{Q}}{2})=e^{-ik_ya_y}e^{i\theta_n(\mathbf{k})} a_n(\mathbf{k}+\frac{\mathbf{Q}}{2}),\mathbf{Q}=(\frac{\pi}{a_x},0),\\
     &a_n(\mathbf{k})=p_na_n(-\mathbf{k}),b_n(\mathbf{k})=p_nb_n(-\mathbf{k}),
\end{aligned}
\label{symm}
\end{equation}
where $\theta_n(\mathbf{k})$ are some model-dependent phase factors and $\theta_n(\mathbf{k})=0$ in the generalized  Kapit-Mueller model~\cite{shen2025}. $p_n=\pm1$ is the  parity of Bloch wave functions under the inversion symmetry.

We still focus on positive semi-definite interactions: $\hat{V}=\sum_{\mathbf{R}}(\hat{S}_{\mathbf{R}})^{\dagger}\hat{S}_{\mathbf{R}}$ where $\hat{S}_{\mathbf{R}}$ are fermion quadratic operators in the form $\hat{S}_{\mathbf{R}}\sim\hat{c}^{\dagger}\hat{c}$. Now, instead of starting from order parameters satisfying QGN, we start from simple interactions and then try to find solvable $\eta$-pairing ground states: $|\psi\rangle=(\hat{P}\hat{\eta}^{\dagger}\hat{P})|\text{vac}\rangle$. Now we do not require the order parameters $\hat{\eta}^{\dagger}$ to have QGN and they could be very complicated in principle.

 We  consider the following  building blocks: $ \hat{S}_{\mathbf{R}}=(\hat{c}^{\dagger}_{A}(\mathbf{R})-\hat{c}^{\dagger}_{B}(\mathbf{R}+\mathbf{a}_y))(\hat{c}_{A}(\mathbf{R})+\hat{c}_{B}(\mathbf{R}+\mathbf{a}_y))$, and the resulting interaction is a simple  density-density attraction: 
 \begin{equation}
     \hat{V}=-2\sum_{\mathbf{R}}\left(2\hat{n}_A(\mathbf{R})\hat{n}_B(\mathbf{R}+\mathbf{a}_y)-\hat{c}^{\dagger}_A(\mathbf{R})\hat{c}_B(\mathbf{R}+\mathbf{a}_y)-\hat{c}^{\dagger}_B(\mathbf{R}+\mathbf{a}_y)\hat{c}_A(\mathbf{R})\right)+2\hat{N},
 \end{equation}
 where $\hat{N}$ is the total density and can be neglected in a particle-number fixed Hilbert space.  To find  the $\hat{P}\hat{\eta}^{\dagger}\hat{P}$ which commutes with all the $\hat{P}\hat{S}_{\mathbf{R}}\hat{P}$, we first consider the projected $\hat{P}\hat{S}_{\mathbf{R}}\hat{P}$ in the energy-band basis:
 \begin{equation}
     \begin{aligned}
\hat{P}\hat{S}_{\mathbf{R}}\hat{P}=\sum_{\mathbf{p},\mathbf{q}}\left( u^*_-(\mathbf{p})u_+(\mathbf{q})\right)e^{-i\mathbf{R}\cdot(\mathbf{p}-\mathbf{q})}\hat{\gamma}^{\dagger}(\mathbf{p})\hat{\gamma}(\mathbf{q}),
\end{aligned}
 \end{equation}
 where the form factor $u^*_-(\mathbf{p})=a^*_n(\mathbf{p})-b_n^*(\mathbf{p})e^{-ip_ya_y},u_+(\mathbf{q})=a_n(\mathbf{q})+b_n(\mathbf{q})e^{iq_ya_y}$. Now we can construct the anti-symmetric order parameter matrix as: $F(\mathbf{p})=\frac{u^*_-(\mathbf{p}+\frac{\mathbf{Q}}{2})}{u_+(\frac{\mathbf{Q}}{2}-\mathbf{p})}=p_n\frac{a_n^*(\mathbf{p}+\frac{\mathbf{}Q}{2})-e^{-ip_ya_y}b_n^*(\mathbf{p}+\frac{\mathbf{}Q}{2})}{a_n(\mathbf{p}-\frac{\mathbf{}Q}{2})+e^{-ip_ya_y}b_n(\mathbf{p}-\frac{\mathbf{Q}}{2})}$. The form factor $ u^*_-(\mathbf{p})u_+(\mathbf{q})$ can be written as: $ u^*_-(\mathbf{p})u_+(\mathbf{q})=F(\mathbf{p}-\frac{\mathbf{Q}}{2})A(\mathbf{p}-\frac{\mathbf{Q}}{2},\frac{\mathbf{Q}}{2}-\mathbf{q})$, where $A(\mathbf{p},\mathbf{q})=u_+(\frac{\mathbf{Q}}{2}-\mathbf{q})u_+(\frac{\mathbf{Q}}{2}-\mathbf{p})$ is symmetric: $A(\mathbf{p},\mathbf{q})=A(\mathbf{q},\mathbf{p})$.

 We can directly verify the $F(\mathbf{p})$ we construct is anti-symmetric (now we have only one flat band, so this is equivalent to $F(\mathbf{p})$ is an odd function):
 \begin{equation}
     \begin{aligned}
         F(-\mathbf{p})&=p_n\frac{a_n^*(\mathbf{p}-\frac{\mathbf{Q}}{2})-b_n^*(\mathbf{p}-\frac{\mathbf{Q}}{2})e^{ip_ya_y}}{a_n(\mathbf{p}+\frac{\mathbf{Q}}{2})+b_n(\mathbf{p}+\frac{\mathbf{Q}}{2})e^{ip_ya_y}}\\
                       &=p_n\frac{e^{ip_ya_y}b_n^*(\mathbf{p}+\frac{\mathbf{Q}}{2})-a_n^*(\mathbf{p}+\frac{\mathbf{Q}}{2})e^{2ip_ya_y}}{e^{ip_ya_y}b_n(\mathbf{p}-\frac{\mathbf{Q}}{2})+a_n(\mathbf{p}-\frac{\mathbf{Q}}{2})e^{2ip_ya_y}}\\
                       &=p_n\frac{e^{-ip_ya_y}b_n^*(\mathbf{p}+\frac{\mathbf{Q}}{2})-a_n^*(\mathbf{p}+\frac{\mathbf{Q}}{2})}{e^{-ip_ya_y}b_n(\mathbf{p}-\frac{\mathbf{Q}}{2})+a_n(\mathbf{p}-\frac{\mathbf{Q}}{2})}\\
                       &=-F(\mathbf{p}),
     \end{aligned}
 \end{equation}
 where we only use the properties in Eq.~\eqref{symm}, which only depend on the magnetic translation and inversion symmetries.

 Hence, with the legitimate  order parameter $\hat{P}\hat{\eta}^{\dagger}\hat{P}=\sum_{\mathbf{p}}F(\mathbf{p})\hat{\gamma}^{\dagger}(\mathbf{p}+\frac{\mathbf{Q}}{2})\hat{\gamma}^{\dagger}(-\mathbf{p}+\frac{\mathbf{Q}}{2})$, 
 the $\eta$-pairing state $|\psi\rangle=(\hat{P}\hat{\eta}^{\dagger}\hat{P})^N|\text{vac}\rangle$ is the exact ground state of the flat band interaction $\sum_{\mathbf{R}}(\hat{\bar{S}}_{\mathbf{R}})^{\dagger}\hat{\bar{S}}_{\mathbf{R}}$, where $\hat{\bar{S}}_{\mathbf{R}}$ is the projected operator $\hat{P}\hat{S}_{\mathbf{R}}\hat{P}$. The only issue left is $\sum_{\mathbf{R}}(\hat{\bar{S}}_{\mathbf{R}})^{\dagger}\hat{\bar{S}}_{\mathbf{R}}$, where $\hat{\bar{S}}_{\mathbf{R}}$ is the projected operator $\hat{P}\hat{S}_{\mathbf{R}}\hat{P}$, is not the same  as $\hat{P}\hat{V}\hat{P}$, and they differ by a quadratic term \cite{han2025}:
 \begin{equation}
     \begin{aligned}
         \hat{P} \hat{H}_{\mathrm{int}, 2} \hat{P}& = \sum_{\boldsymbol{R}} \hat{P}\hat{S}_{\boldsymbol{R}}^{\dagger} \hat{S}_{\boldsymbol{R}}\hat{P}-\hat{P}\hat{S}_{\boldsymbol{R}}^{\dagger} \hat{P} \hat{S}_{\boldsymbol{R}} \hat{P}\\  
         &=\frac{1}{\mathrm{S}} \sum_{\substack{\boldsymbol{p q}, \boldsymbol{p}^{\prime} \boldsymbol{q}^{\prime} \\ n m k l}} \delta_{\boldsymbol{p}^{\prime}-\boldsymbol{q}^{\prime}, \boldsymbol{q}-\boldsymbol{p}} S_{m n}^{*}(\boldsymbol{q}, \boldsymbol{p}) S_{k l}\left(\boldsymbol{p}^{\prime}, \boldsymbol{q}^{\prime}\right) \hat{P}\hat{\gamma}_{\mathbf{p},n}^{\dagger} \hat{\gamma}_{\boldsymbol{q}, m}(1-\hat{P}) \hat{\gamma}_{\boldsymbol{p}^{\prime}, k}^{\dagger} \hat{\gamma}_{\boldsymbol{q}^{\prime}, l} \hat{P} \\ 
         &=\frac{1}{\mathrm{S}} \sum_{\substack{\boldsymbol{p q} \\ n l \leq \mathrm{N}_{\mathrm{flat}}, m>\mathrm{N}_{\mathrm{flat}}}}  S_{n m}^{\dagger}(\boldsymbol{q}, \boldsymbol{p}) S_{m l}(\boldsymbol{q}, \boldsymbol{p}) \hat{\gamma}_{\boldsymbol{p}, n}^{\dagger} \hat{\gamma}_{\boldsymbol{p}, l},
     \end{aligned}
 \end{equation}
 which can be canceled by the projection of a quadratic term :
 \begin{equation}
     \begin{aligned}
         \hat{P}\hat{V}_2\hat{P}&=-\alpha\hat{P}\left[\sum_{\mathbf{R}}(\hat{c}^{\dagger}_{A}(\mathbf{R})+\hat{c}^{\dagger}_{B}(\mathbf{R}+\mathbf{a}_y))(\hat{c}_{A}(\mathbf{R})+\hat{c}_{B}(\mathbf{R}+\mathbf{a}_y))\right]\hat{P}\\
         &=-\alpha\hat{P}\left[\sum_{\mathbf{R}}\hat{c}^{\dagger}_{A}(\mathbf{R})\hat{c}_{B}(\mathbf{R}+\mathbf{a}_y)+\hat{c}^{\dagger}_{B}(\mathbf{R}+\mathbf{a}_y)\hat{c}_{A}(\mathbf{R})+\hat{N}\right]\hat{P},
     \end{aligned}
 \end{equation}
  where the coefficient $\alpha=\frac{1}{S}\sum_{\mathbf{p}}|b_n^*(\mathbf{p})+a^*_n(\mathbf{p})e^{ip_ya_y}|^2$. As a result, the total interaction is $\hat{V}_{\text{tot}}=\hat{V}+\hat{V}_2=-4\sum_{\mathbf{R}}\hat{n}_{A}(\mathbf{R})\hat{n}_{B}(\mathbf{R}+\mathbf{a}_y)+(2-\alpha)\left[\sum_{\mathbf{R}}\hat{c}^{\dagger}_{A}(\mathbf{R})\hat{c}_{B}(\mathbf{R}+\mathbf{a}_y)+\hat{c}^{\dagger}_{B}(\mathbf{R}+\mathbf{a}_y)\hat{c}_{A}(\mathbf{R}))+\hat{N}\right]$. We want to emphasize again that the quadratic term in $\hat{V}_{\text{tot}}$ is to make the projected $\hat{P}\hat{V}_{\text{tot}}\hat{P}$ only contains four fermion interaction terms.
 \subsection{Chern number}
 We can show the Chern number of the above FFLO state is nonzero if the ground state is a gapped superconductor, and we can argue that there is indeed a single-particle gap of our FFLO state. First, in order to study the Chern number, it is more convenient to consider the BCS state, which is also a ground state:
 \begin{equation}
     |\alpha\rangle=\sum^{+\infty}_{N=0}e^{-\alpha N}\frac{(\tilde{\eta}^{\dagger}/2)^N}{N!}|\text{vac}\rangle=\Pi^{\prime}_{\mathbf{p}}\left[1+e^{-\alpha}F(\mathbf{p})\hat{\gamma}^{\dagger}(\mathbf{p}+\frac{\mathbf{Q}}{2})\hat{\gamma}^{\dagger}(-\mathbf{p}+\frac{\mathbf{Q}}{2})\right]|\text{vac}\rangle,
 \end{equation}
 where $\alpha$ is a complex parameter conjugate with the particle number $N$, and the $\tilde{\eta}^{\dagger}=P\hat{\eta}^{\dagger}P=\sum_{\mathbf{p}}F(\mathbf{p})\hat{\gamma}^{\dagger}(\mathbf{p}+\frac{\mathbf{Q}}{2})\hat{\gamma}^{\dagger}(-\mathbf{p}+\frac{\mathbf{Q}}{2})$. We use $\frac{\tilde{\eta}^{\dagger}}{2}$ in the BCS state construction to avoid double counting of the momentum and the $\Pi^{\prime}_{\mathbf{p}}$ in the final equality only includes half of the Brillouin zone.

 The BCS state $|\alpha\rangle$ can be viewed as a state with a fully occupied BdG band defined by: $\chi^{\dagger}_{\mathbf{p}}=\frac{e^{-\alpha}F(\mathbf{p})\gamma^{\dagger}(\mathbf{p}+\frac{Q}{2})+\hat{\gamma}(-\mathbf{p}+\frac{\mathbf{Q}}{2})}{\sqrt{1+e^{-2\Re[\alpha]}|F(\mathbf{p})|^2}}$ and $\chi^{\dagger}_{\mathbf{p}}|\alpha\rangle=0$. $\Re[\alpha]$ means the real part of the complex parameter $\alpha$. Now we can obtain the BdG wave function of this BCS state in the sublattice representation:
 \begin{equation}
     \begin{aligned}
         \chi^{\dagger}_{\mathbf{p}}&=\left(\begin{array}{ll}\hat{\gamma}_{\boldsymbol{p}+\frac{\mathbf{Q}}{2} }^{\dagger} & \hat{\gamma}_{-\boldsymbol{p} +\frac{\mathbf{Q}}{2}}\end{array}\right)\binom{e^{-\alpha} u^*_{-}(\boldsymbol{p}+\frac{\mathbf{Q}}{2}) }{u_{+}(-\boldsymbol{p}+\frac{\mathbf{Q}}{2})}\frac{1}{\sqrt{|u_{+}(-\boldsymbol{p}+\frac{\mathbf{Q}}{2})|^2+e^{-2\Re[\alpha]} |u_{-}(\boldsymbol{p}+\frac{\mathbf{Q}}{2})|^2}}\\
         &=\left(\begin{array}{ll}\hat{c}_{\boldsymbol{p}+\frac{\mathbf{Q}}{2},\mu }^{\dagger} & \hat{c}_{-\boldsymbol{p} +\frac{\mathbf{Q}}{2},\nu}\end{array}\right)\binom{U_{\mu1}(\mathbf{p}+\frac{\mathbf{Q}}{2})e^{-\alpha} u^*_{-}(\boldsymbol{p}+\frac{\mathbf{Q}}{2}) }{(U_{\nu1}(-\mathbf{p}+\frac{\mathbf{Q}}{2}))^*u_{+}(-\boldsymbol{p}+\frac{\mathbf{Q}}{2})}\frac{1}{\sqrt{|u_{+}(-\boldsymbol{p}+\frac{\mathbf{Q}}{2})|^2+e^{-2\Re[\alpha]} |u_{-}(\boldsymbol{p}+\frac{\mathbf{Q}}{2})|^2}}\\
         &=\left(\begin{array}{ll}\hat{c}_{\boldsymbol{p}+\frac{\mathbf{Q}}{2},\mu }^{\dagger} & \hat{c}_{-\boldsymbol{p} +\frac{\mathbf{Q}}{2},\nu}\end{array}\right)w(\mathbf{p}),
     \end{aligned}
 \end{equation}
 where $\mu,\nu\in \{A,B\}$ are the sublattice indices and $U_{A1}(\mathbf{k})=a_n(\mathbf{k}),U_{B1}(\mathbf{k})=b_n(\mathbf{k})$. $w(\mathbf{p})$ is a four-component wave function since it contains both the sublattice and BdG indices.

 Now we argue that the FFLO state is generally gapped except for some coincidental nodes, and the argument is similar to the chiral superconductor in \cite{han2025}. This is implied by the wave function $w(\mathbf{p})$, which is continuous and well-defined (since the denominator can never be zero) across the Brillouin zone. Then the Chern number of the FFLO state is:
 \begin{equation}
     C=C^{\mathrm{BdG}}-C_{>},
 \end{equation}
 where $C^{\mathrm{BdG}}$ is the fully occupied BdG band defined above, and $C_{>}$ is the Chern number of the empty upper band. We have to include $C_{>}$ as the upper band is fully occupied by holes in the BdG representation $\left(\begin{array}{ll}\hat{c}_{\boldsymbol{p}+\frac{\mathbf{Q}}{2},\mu }^{\dagger} & \hat{c}_{-\boldsymbol{p} +\frac{\mathbf{Q}}{2},\nu}\end{array}\right)$. We note that the BdG wave function $w(\mathbf{p})$ is periodic across the Brillouin zone, so $C^{\mathrm{BdG}}=0$. As a result, the total Chern number $|C|=|C_{>}|=1$, which means our FFLO is a topological superconductor with chiral Majorana edge modes.

\end{document}